\documentclass[twoside,twocolumn,9pt]{article}
\usepackage{extsizes}
\usepackage[super,sort&compress,comma]{natbib} 
\usepackage[version=3]{mhchem}
\usepackage[left=1.5cm, right=1.5cm, top=1.785cm, bottom=2.0cm]{geometry}
\usepackage{balance}
\usepackage{mathptmx}
\usepackage{sectsty}
\usepackage{graphicx} 
\usepackage{lastpage}
\usepackage[format=plain,justification=justified,singlelinecheck=false,font={stretch=1.125,small,sf},labelfont=bf,labelsep=space]{caption}
\usepackage{float}
\usepackage{fancyhdr}
\usepackage{fnpos}
\usepackage[english]{babel}
\addto{\captionsenglish}{%
}
\usepackage{array}
\usepackage{droidsans}
\usepackage{charter}
\usepackage[T1]{fontenc}
\usepackage[usenames,dvipsnames]{xcolor}
\usepackage{setspace}
\usepackage[compact]{titlesec}
\usepackage{hyperref}
\usepackage{epstopdf}%This line makes .eps figures into .pdf - please comment out if not required.

\definecolor{cream}{RGB}{222,217,201}

\begin{document}

\pagestyle{fancy}
\thispagestyle{plain}
\fancypagestyle{plain}{
%%%HEADER%%%
\renewcommand{\headrulewidth}{0pt}
}
%%%END OF HEADER%%%

%%%PAGE SETUP - Please do not change any commands within this section%%%
\makeFNbottom
\makeatletter
\renewcommand\LARGE{\@setfontsize\LARGE{15pt}{17}}
\renewcommand\Large{\@setfontsize\Large{12pt}{14}}
\renewcommand\large{\@setfontsize\large{10pt}{12}}
\renewcommand\footnotesize{\@setfontsize\footnotesize{7pt}{10}}
\makeatother

\renewcommand{\thefootnote}{\fnsymbol{footnote}}
\renewcommand\footnoterule{\vspace*{1pt}% 
\color{cream}\hrule width 3.5in height 0.4pt \color{black}\vspace*{5pt}} 
\setcounter{secnumdepth}{5}

\makeatletter 
\renewcommand\@biblabel[1]{#1}            
\renewcommand\@makefntext[1]% 
{\noindent\makebox[0pt][r]{\@thefnmark\,}#1}
\makeatother 
\renewcommand{\figurename}{\small{Fig.}~}
\sectionfont{\sffamily\Large}
\subsectionfont{\normalsize}
\subsubsectionfont{\bf}
\setstretch{1.125} %In particular, please do not alter this line.
\setlength{\skip\footins}{0.8cm}
\setlength{\footnotesep}{0.25cm}
\setlength{\jot}{10pt}
\titlespacing*{\section}{0pt}{4pt}{4pt}
\titlespacing*{\subsection}{0pt}{15pt}{1pt}
%%%END OF PAGE SETUP%%%

%%%FOOTER%%%
\fancyfoot{}
\fancyfoot[LO,RE]{\vspace{-7.1pt}\includegraphics[height=9pt]{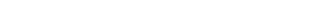}}
\fancyfoot[CO]{\vspace{-7.1pt}\hspace{11.9cm}\includegraphics{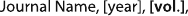}}
\fancyfoot[CE]{\vspace{-7.2pt}\hspace{-13.2cm}\includegraphics{head_foot/RF}}
\fancyfoot[RO]{\footnotesize{\sffamily{1--\pageref{LastPage} ~\textbar  \hspace{2pt}\thepage}}}
\fancyfoot[LE]{\footnotesize{\sffamily{\thepage~\textbar\hspace{4.65cm} 1--\pageref{LastPage}}}}
\fancyhead{}
\renewcommand{\headrulewidth}{0pt} 
\renewcommand{\footrulewidth}{0pt}
\setlength{\arrayrulewidth}{1pt}
\setlength{\columnsep}{6.5mm}
\setlength\bibsep{1pt}
%%%END OF FOOTER%%%

%%%FIGURE SETUP - please do not change any commands within this section%%%
\makeatletter 
\newlength{\figrulesep} 
\setlength{\figrulesep}{0.5\textfloatsep} 

\newcommand{\topfigrule}{\vspace*{-1pt}% 
\noindent{\color{cream}\rule[-\figrulesep]{\columnwidth}{1.5pt}} }

\newcommand{\botfigrule}{\vspace*{-2pt}% 
\noindent{\color{cream}\rule[\figrulesep]{\columnwidth}{1.5pt}} }

\newcommand{\dblfigrule}{\vspace*{-1pt}% 
\noindent{\color{cream}\rule[-\figrulesep]{\textwidth}{1.5pt}} }

\makeatother
%%%END OF FIGURE SETUP%%%

%%%TITLE, AUTHORS AND ABSTRACT%%%
\twocolumn[
  \begin{@twocolumnfalse}
{\includegraphics[height=30pt]{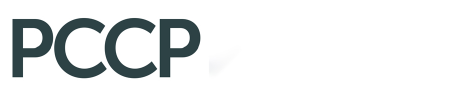}\hfill\raisebox{0pt}[0pt][0pt]{\includegraphics[height=55pt]{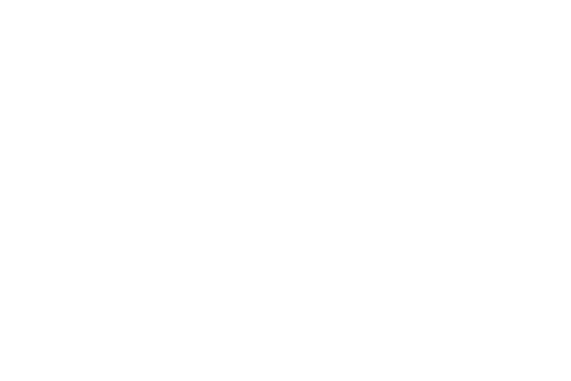}}\\[1ex]
\includegraphics[width=18.5cm]{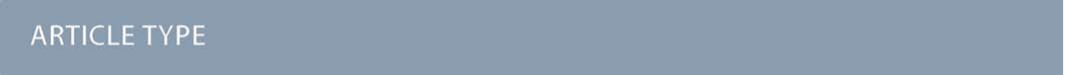}}\par
\vspace{1em}
\sffamily
\begin{tabular}{m{4.5cm} p{13.5cm} }

\includegraphics{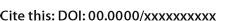} & \noindent\LARGE{\textbf{Neural network ensemble for computing cross sections for rotational transitions in H$_{2}$O + H$_{2}$O collisions}} \\%Article title goes here instead of the text "This is the title"
\vspace{0.3cm} & \vspace{0.3cm} \\

 & \noindent\large{Bikramaditya Mandal,\textit{$^{a}$} Dmitri Babikov,\textit{$^{b}$} Phillip C. Stancil,\textit{$^{c}$} Robert C. Forrey,\textit{$^{d}$} Roman V. Krems,\textit{$^{e}$} and Naduvalath Balakrishnan$^{\ast}$\textit{$^{a}$}} \\%Author names go here instead of "Full name", etc.

\includegraphics{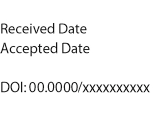} & \noindent\normalsize{Water (H$_2$O) is one of the most abundant molecules in the universe and is found in a wide variety of astrophysical environments. Rotational transitions in H$_2$O + H$_2$O collisions are important in modeling environments rich in water molecules but they are computationally intractable using quantum mechanical methods. Here, we present a machine learning (ML) tool using an ensemble of neural networks (NNs) to predict cross sections to construct a database of rate coefficients for rotationally inelastic transitions in collisions of complex molecules such as water. The proposed methodology utilizes data computed with a mixed quantum-classical theory (MQCT). We illustrate that efficient ML models using NN can be built to accurately interpolate in the space of 12 quantum numbers for rotational transitions in two asymmetric top molecules, spanning both initial and final states. We examine various architectures of data corresponding to each collision energy, symmetry of water molecule, and excitation/de-excitation rotational transitions, and optimize the training/validation data sets. Using only about 10\% of the computed data for training, the NNs predict cross sections of state-to-state rotational transitions of H$_{2}$O + H$_{2}$O collision with average relative root mean square error of 0.409. Thermally averaged cross sections, computed using the predicted state-to-state cross sections ($\sim$90\%) and the data used for training and validation ($\sim$10\%) were compared against those obtained entirely from MQCT calculations. The agreement is found to be excellent with an average percent deviation of about $\sim$13.5\%. The methodology is robust, and thus, applicable to other complex  molecular systems.} \\%The abstrast goes here instead of the text "The abstract should be..."

\end{tabular}

 \end{@twocolumnfalse} \vspace{0.6cm}

  ]
%%%END OF TITLE, AUTHORS AND ABSTRACT%%%

%%%FONT SETUP - please do not change any commands within this section
\renewcommand*\rmdefault{bch}\normalfont\upshape
\rmfamily
\section*{}
\vspace{-1cm}

%%%FOOTNOTES%%%

\footnotetext{\textit{$^{a}$~Department of Chemistry and Biochemistry, University of Nevada, Las Vegas, Nevada 89154, USA}}
\footnotetext{\textit{$^{b}$~Department of Chemistry, Marquette University, Milwaukee, WI 53233, USA}}
\footnotetext{\textit{$^{c}$~Department of Physics and Astronomy and the Center for Simulational Physics, University of Georgia, Athens, GA 30602}}
\footnotetext{\textit{$^{d}$~Department of Physics, Penn State University, Berks Campus, Reading, PA 19610}}
\footnotetext{\textit{$^{e}$~Department of Chemistry, University of British Columbia, Vancouver, BC Canada, V6T 1Z1}}
\footnotetext{\textit{$^{*}$~E-mail: naduvala@unlv.nevada.edu}}

%Please use \dag to cite the ESI in the main text of the article.
%If you article does not have ESI please remove the the \dag symbol from the title and the footnotetext below.
\footnotetext{\dag~Supplementary Information available: [Neural network ensemble for computing cross sections for rotational transitions in H$_{2}$O + H$_{2}$O collisions. Additional supporting material on ML models for different datasets of training, validation and testing and comparison of model predictions against original MQCT data]. See DOI: 10.1039/cXCP00000x/}
%additional addresses can be cited as above using the lower-case letters, c, d, e... If all authors are from the same address, no letter is required

%\footnotetext{\ddag~Additional footnotes to the title and authors can be included \textit{e.g.}\ `Present address:' or `These authors contributed equally to this work' as above using the symbols: \ddag, \textsection, and \P. Please place the appropriate symbol next to the author's name and include a \texttt{\textbackslash footnotetext} entry in the the correct place in the list.}

%%%END OF FOOTNOTES%%%

%%%MAIN TEXT%%%%
\section{Introduction}
Several state-of-the-art observatories, such as Atacama Large Millimeter Array (ALMA) radio telescope, and other space-exploration telescopes, including the Spitzer space telescope and, most recently, the James Webb space telescope (JWST), have been deployed to collect spectroscopic data. The spectra obtained from these telescopes show  signatures of water molecules with isotopic constitutions, namely H$_{2}$O and HDO. Observation and analysis of water isotopologues in cometary comae, in relation to their abundances on Earth and other solar system bodies, can yield valuable insights into the early history of Earth and, by extension, the solar system. Water is also found in a large variety of astrophysical environments. For example, water is detected in cometary comae~\cite{bockelee2015cometary, dones2015origin}, cold interstellar molecular clouds~\cite{van2021water}, stellar photospheres and circumstellar envelopes~\cite{van2021water}, and atmospheres of icy planets~\cite{hartogh2011direct}. Water represents the major reservoir of oxygen and, thus, controls the chemistry of many species in the gas phase and also on  grain surfaces~\cite{roueff2013molecular, van2021water}. In star-forming regions, water emission dominates the process of gas cooling~\cite{van2011water}. This process creates the brightest line in the radio frequency or maser-like radiation in the GHz range~\cite{humphreys2007submillimeter}, which carries information about physical conditions in these environments. The low temperature in prestellar core conditions leads to freezing of most volatile compounds  onto surfaces of grains~\cite{mandt2015constraints}. The variability of the D/H ratio in different molecules, particularly in water, can yield essential details on its formation conditions in our solar system.
		
Atmospheres of icy planets and their moons, such as Jovian moons, are known to have an anisotropic distribution of water vapor, affecting the properties of the observed water line. While most of such atmosphere is collision-less, the sub-solar point supports intense sublimation and photo-induced desorption, which results in a distribution that is not in a local thermodynamic equilibrium (non-LTE) and driven by molecular collisions, such as excitation, and de-excitation of H$_{2}$O. Interpreting these and many other observations requires numerical modeling and relies on the knowledge of precise excitation and quenching schemes for ortho- and para-H$_{2}$O. Large uncertainties of rate coefficients for these transitions can affect the predictions of astrophysical models by orders of magnitude~\cite{loreau2018scattering, khalifa2020rotational}. To characterize emission as a function of coma radius, modeling with radiation transfer codes, such as RADEX \cite{van2007computer}, LIME \cite{brinch2010lime}, or MOLPOP \cite{ramos2018molpop}, is necessary, which in turn requires collision rate coefficients as input. To understand and correlate with the observed rotational spectra from ALMA or JWST missions, one requires state-to-state collisional rate coefficients for the rotational excitation and quenching processes. These rate coefficients are difficult to calculate for complex collision systems such as H$_{2}$O + H$_{2}$O and HDO + H$_{2}$O. Besides these, some of the much needed rate coefficients for inelastic collisions for astrophysical models include CH$_{3}$OH + CO, H$_{2}$O + HCN, H$_{2}$CO + CO, and H$_{2}$O + CH$_{3}$OH~\cite{biver2022observations}. 
		
Databases such as BASECOL \cite{dubernet2024basecol2023} and LAMDA \cite{van2020leiden} have been developed  to simplify the process of obtaining rate coefficients for  different molecular systems. The rate coefficients can be computed by a quantum mechanical treatment of the collision problem as implemented in a few codes available to the scientific community, such as MOLSCAT \cite{hutson2020molscat,hutson2019molscat}, HIBRIDON \cite{alexander2023hibridon}, and TwoBC~\cite{Krems2006} or using a mixed quantum/classical theory (MQCT)~\cite{mandal2024mqct}, when quantum calculations are not practical.
		
The inelastic collisions of H$_{2}$O with H$_{2}$O are almost impossible to study using full quantum methods because a large number of combined states of both water molecules needs to be considered, and a water molecule has a dense spectrum of quantum states~\cite{agg1991infinite, boursier2020new, buffa2000h2o}. However, significant progress has been made recently by Mandal et al. to study rotationally inelastic collisions of two water molecules using the MQCT~\cite{mandal2024rotational, mandal2023rate, mandal2023improved}. MQCT has proven its ability to produce accurate result with computational efficiency for inelastic collisions of several  molecular systems~\cite{mandal2024rotational, mandal2023rate, mandal2018calculations, mandal2020adiabatic, mandal2021development, mandal2022mixed, mandal2023adiabatic, mandal2023improved, mandal2024mqct, mandal2024rotational, bostan2023description, boursier2020new, bostan2024mixed, joy2023mixed, joy2024mixed, joy2024rate, bostan2024mixedCOCO, semenov2020mqct}. In this approach, the relative translational motion of collision partners is treated classically using the mean-field trajectories method, while rotations and vibrations ($i.e.$, internal degrees of freedom of the colliding molecules) are treated quantum mechanically~\cite{mandal2024mqct}. In the current implementation of the MQCT method for  H$_{2}$O + H$_{2}$O collisions, the H$_2$O molecules are treated as rigid rotors~\cite{mandal2024mqct}. 
		
Using the MQCT methodology, a database of thermally averaged cross sections (TACS) (averaged over a thermal population of rotational levels of the partner H$_2$O molecule) was first published by Mandal \& Babikov~\cite{mandal2023rate} followed by a database of thermal rate coefficients~\cite{mandal2023improved}. This database of TACS and the rate coefficients contained 231 transitions in para-H$_{2}$O  and 210 transitions in ortho-H$_{2}$O (both treated as the target molecule) in the temperature range $5\le T\le1000$ K. In a subsequent study, a database of both rotational temperature ($T_{\rm{rot}}$) and kinetic temperature ($T_{\rm{kin}}$) dependent rate coefficients was built to model non-LTE environments using RADEX for H$_{2}$O + H$_{2}$O collisions~\cite{mandal2024rotational}.
		
While MQCT can be applied to collisions of two water molecules, the computational complexity remains challenging. As elaborated by Mandal \& Babikov~\cite{mandal2023rate}, the computation  involves evaluation of matrix elements of the interaction potential in a basis of rotational wave functions of the water molecules that are required for the simulation of mixed quantum/classical trajectories. In the prior works, the computation of these matrices alone required about $\sim$2.7 M CPU hours in the HPC facility Raj at Marquette University (AMD Rome 2 GHz processors, memory 512 GB). Additionally, the total cost of the scattering calculations (trajectory simulations) for six collision energies was about 5.25 M CPU hours using the same HPC facility. Altogether, the cost of the MQCT calculations of rate-coefficients for H$_{2}$O + H$_{2}$O collisions was nearly 8 million CPU hours. More importantly, several months of human work were needed to manage the ongoing simulations, and carry out numerous post-processing analyses to convert state-to-state  cross sections to the rate coefficients to be deposited into the databases. While  significant speedup in the computation of the relevant coupling matrices has been achieved recently, the trajectory simulations remain computationally demanding.
				
The challenge of such massive computational tasks is two-fold. First, computing the TACS for 231 transitions in para- and 210 transitions in ortho-H$_{2}$O target required over a million cross sections for individual state-to-state transition at each collision energy. Independent calculations for a total of 3268 initial states combining both target and quencher molecule needed to be completed to build the database. Secondly, each of these simulations for individual initial states and collision energies required computation of four large interaction potential matrices, each containing about 1.35 million coupling elements. With these two factors, building an extended database for such a complex system using direct calculations is often not practical. Machine learning may instead prove useful for constructing such a complex database.
		
Machine learning (ML) has found widespread applications in recent years in many areas of physics and chemistry, including condensed matter physics~\cite{moustafa2023hundreds, rahaman2023machine, ciarella2023dynamics}, nuclear physics~\cite{ma2023simple, nobre2023novel, mallick2023deep}, astronomy~\cite{fang2023automatic, jasinski2020machine, lochner2023unique}, particle physics~\cite{ilten2023modeling, butter2023machine, eller2023flexible}, quantum many-body physics~\cite{anstine2023machine, miles2023machine, zhang2023transformer}, cosmology~\cite{lemos2023robust, de2023machine, gomez2023neural} as well as fitting of multi-dimensional potential energy surfaces (PESs) from electronic structure calculations. Permutationally invariant polynomials (PIPs) combined with neural networks (NNs) by Bowman, Guo, and others \cite{braams2009permutationally, qu2018permutationally, houston2024no, jiang2013permutation, li2013permutation, jiang2014permutation, xie2018permutation, jiang2016potential, biswas2020machine, biswas2023artificial, biswas2023improved, kushwaha2023benchmarking} and Gaussian process regression (GPR) by Krems and coworkers~\cite{dai2023neural, dai2022quantum, asnaashari2021gradient, sugisawa2020gaussian, dai2020interpolation, krems2019bayesian}, have been widely adopted in building PESs of complex molecular systems. ML has also been used to predict rate coefficients for inelastic collisions  of diatomic molecules for astrophysical modeling from a smaller set of available data and improve the accuracy of approximate quantum scattering calculations~\cite{jasinski2020machine, meuwly2021machine, bossion2024machine, arnold2020machine, krems2019bayesian, mihalik2025accurate, wang2025multilayer}. Quantum machine learning is also being actively explored~\cite{torabian2023compositional, dawid2022modern, kairon2025equivalence, zhang2020recent, cerezo2022challenges, ciliberto2018quantum, schuld2015introduction}.
		
To our knowledge, ML has so far not been applied to develop a complex database of rate coefficients for collisions involving two triatomic molecules. In this work, our goal is to reduce the computational effort needed to build databases for complex colliding partners, such as H$_{2}$O + H$_{2}$O, by implementing and incorporating ML algorithms into this process so that more such databases can be produced, and made available to the modeling community. For this purpose, we make use of previously computed cross sections for individual state-to-state transitions for collisions of two water molecules as a benchmark, and for training machine learning models. The goal is to use the available data to explore if it is feasible to construct a reliable model for efficient interpolation in the large space of quantum states of two triatomic molecules.
		
The paper is organized as follows: Section \ref{sec:methods} briefly discusses the theory to compute thermally-averaged rate coefficients for H$_{2}$O + H$_{2}$O collisions, data pre-processing and architecture of the machine learning models employed here. In section \ref{sec:results} we discuss results obtained from the ML models using NN and compare them against MQCT data not used in training the models. A summary of our findings is given in section \ref{sec:conclusion}.

\section{Methods}
\label{sec:methods}
\subsection{Thermally-averaged cross sections (TACS)}
The process of building a database of rotationally inelastic rate coefficients for H$_{2}$O + H$_{2}$O collisions by computing  thermally averaged rate coefficients, $k_{n_{1}^{}\to n_{1}'}(T_{\rm{rot}}, T_{\rm{kin}})$, is explained in detail by Mandal et al.~\cite{mandal2024rotational}. Only relevant equations to compute the TACSs are provided below.
				
Thermal rate coefficients for state-to-state transitions for a given kinetic temperature $T_{\rm{kin}}$ are computed by averaging the corresponding cross sections over a Maxwell-Boltzmann distribution of relative velocities for all relevant collision energies, $E_{c}$ as follows:
\begin{equation}
	\begin{aligned}
		k_{n_{1}^{}n_{2}^{}\to n_{1}'n_{2}'}(T_{\rm{kin}})=&\frac{v_{\rm{ave}}(T_{\rm{kin}})}{(k_{\rm{B}}T_{\rm{kin}})^{2}}\times\\
		&\int\limits_{E_{c}=0}^{\infty}E_{c} ~\sigma_{n_{1}^{}n_{2}^{}\to n_{1}'n_{2}'}(E_{c})~e^{-\frac{E_{c}}{k_{\rm{B}}T_{\rm{kin}}}}dE_{c},
	\end{aligned}
	\label{maxwell_blotzmann_eqn}
\end{equation}
where, $k_{\rm{B}}$ is the Boltzmann constant, $v_{\rm{ave}}(T_{\rm{kin}})=\sqrt{8k_{\rm{B}}T_{\rm{kin}}/\pi\mu}$ is the average collision velocity, $\mu$ is the reduced mass of the collision complex, and the subscripts $n_{1}^{}n_{2}^{}$ and $n_{1}'n_{2}'$ indicate the initial and final states, respectively. Each $n$ is a composite index that represents a full set of quantum numbers for one molecule. For example, for water molecules, $n$ denotes $j_{k_{A}^{}k_{C}^{}}$, where $j$ is the rotational quantum number and $k_A$ and $k_C$ are the projections of $j$ along the axis of the largest and smallest moment of inertia, $I_A$ and $I_C$, respectively . Further, for para-H$_2$O (nuclear spins of two H atoms are anti-parallel), $k_A+k_C=$ is even and for ortho-H$_2$O (nuclear spins of two H atoms are parallel), $k_A+k_C=$ is odd. Since our focus is on the target H$_{2}$O molecule, we compute the rate coefficients for water molecule by summing over all final states and averaging over all initial states of its collision partner (quencher):
\begin{equation}
	k_{n_{1}^{}\to n_{1}'}(T_{\rm{rot}}, T_{\rm{kin}})=\sum_{n_{2}}^{}w_{n_{2}}(T_{\rm{rot}})\sum_{n_{2}'}^{}k_{n_{1}^{}n_{2}^{}\to n_{1}'n_{2}'}(T_{\rm{kin}}).
	\label{rate_mol1}
\end{equation}
In  eq. (\ref{rate_mol1}), the thermal populations or weights $w_{n_{2}^{}}(T_{\rm{rot}})$ of the initial states of the quencher are defined as
\begin{equation}
	w_{n_{2}^{}}(T_{\rm{rot}})=\frac{(2j_{2}^{}+1)e^{-\frac{E_{2}^{}}{k_{\rm{B}}T_{\rm{rot}}}}}{Q_{2}^{}(T_{\rm{rot}})},
	\label{wgt_eq}
\end{equation}
where, $E_{2}^{}$ represents the energies of the rotational states $n_{2}^{}$ of the quencher. The denominator $Q_{2}^{}(T_{\rm{rot}})$ in eq. (\ref{wgt_eq}) is the rotational partition function of the quencher given by
\begin{equation}
	Q_{2}^{}(T_{\rm{rot}})=\sum_{n_{2}^{}}^{}(2j_{2}^{}+1)e^{-\frac{E_{2}^{}}{k_{\rm{B}}T_{\rm{rot}}}}.
	\label{partition_function}
\end{equation}
For more details of the computation of $Q_{2}$, see Ref~\cite{mandal2024rotational}.
		
Computation of the state-to-state rate coefficients in eq. (\ref{maxwell_blotzmann_eqn}) reaches practical limitation for complex systems, like H$_{2}$O + H$_{2}$O, due to the enormous numbers of individual state-to-state transitions $n_{1}^{}n_{2}^{}\to n_{1}'n_{2}'$. In the work reported by Mandal \& Babikov~\cite{mandal2023rate},  231 para-para and 210 ortho-ortho transitions were computed for the target H$_{2}$O, considering a maximum value of $j_{1}=7$. This required a rotational basis set with 38 states each for the para- and ortho-isomers of the quencher H$_{2}$O molecule for which a maximum value of $j_{2}=10$ was adopted. This led to over a million individual state-to-state transitions $n_{1}^{}n_{2}^{}\to n_{1}'n_{2}'$, considering all the initial states of the molecular system as elaborated  in the introduction. Ideally, these calculations  need to be done on a grid of collision energy dense enough to perform the integral over the collision energy $E_c$, as shown in eq. (\ref{maxwell_blotzmann_eqn}). The human effort needed to manually check millions of individual transitions and implement them in eq. (\ref{maxwell_blotzmann_eqn}) is very labor intensive.
		
To tackle this challenge, an alternative approach was introduced by exchanging the order of integration in eq. (\ref{maxwell_blotzmann_eqn}) with the summation in eq. (\ref{rate_mol1}) as
\begin{equation}
	\begin{aligned}
		k_{n_{1}^{}\to n_{1}'}(T_{\rm{rot}}, T_{\rm{kin}})=&\frac{v_{\rm{ave}}(T_{\rm{kin}})}{(k_{\rm{B}}T_{\rm{kin}})^{2}}\times\\
		&\int\limits_{E_{c}=0}^{\infty} ~\sigma_{n_{1}^{}\to n_{1}'}(E_{c},T_{\rm{rot}})~e^{-\frac{E_{c}}{k_{\rm{B}}T_{\rm{kin}}}}E_{c}dE_{c}.
	\end{aligned}
	\label{TACS_into_rate}
\end{equation}
In  eq. (\ref{TACS_into_rate}), since the summation over the states of the quencher H$_{2}$O molecule is now carried out before the integration over collision energy ($E_{c}$),  a thermally averaged cross section  for the transition $n_{1}^{}\to n_{1}'$ of the target H$_{2}$O molecule is introduced:
\begin{equation}
	\sigma_{n_{1}^{}\to n_{1}'}(E_{c},T_{\rm{rot}}) =\sum_{n_{2}^{}}^{}w_{n_{2}^{}}^{}(T_{\rm{rot}})\sum_{n_{2}'}^{}\sigma_{n_{1}^{}n_{2}^{}\to n_{1}'n_{2}'}(E_{c}).
	\label{TACS_eq}
\end{equation}
		
The TACSs are computed as  follows: first, all the individual state-to-state transition cross sections are summed over the final states of the quencher H$_{2}$O molecule, and then the resulting sums are averaged over the initial states of the quencher H$_{2}$O molecule for a given value of rotational temperature $T_{\rm{rot}}$ and collision energy $E_{c}$. Since the number of rotational transitions between the states of a target molecule is relatively small, it is much easier to check the behavior of all TACS before they are integrated over the collision energy in Eq. (\ref{TACS_into_rate}). As previously mentioned, in the work of Mandal \& Babikov~\cite{mandal2023rate}, the number of rotational transitions in para- and ortho-water considering only de-excitation processes were 231 and 210, respectively,  and the number of collision energies were six, making them  easier to manually check and ensure proper behavior.
		
The computed TACS for those six collision energies can then be used for analytical fits to compute kinetic temperature dependent rate coefficients as described in detail by Mandal et al.~\cite{mandal2024rotational}. However, in this work, the computed TACS are the main goal; therefore, the details of computing rate coefficients using anlytical expressions are not discussed here. These 231 transitions in para- and 210 transitions in ortho-H$_{2}$O for the target molecule are used in this work as a benchmark of the ML predictions.
		
\subsection{Details of machine-learning method}
The TACS described in the previous section require the individual state-to-state transition cross sections considering initial and final states of the target as well as the quencher H$_{2}$O molecules. As stated in the introduction, the goal of the present study is to effectively reduce computational cost of the MQCT calculations to evaluate these state-to-state transition cross sections. This is achieved by the  methodology described in the ensuing sections.
		
\subsubsection{Data analysis and pre-processing for machine learning}
We begin by analyzing the available data for both excitation and de-excitation of the target H$_2$O molecule. Previous studies by Mandal \& Babikov~\cite{mandal2023rate} showed that the dependencies of cross sections on the energy difference between initial and final states of the colliding partners, given by $\Delta E=E_{\rm{initial}} - E_{\rm{final}}$, depicts a single-exponential decay near $\Delta E=0$ regime, and a double-exponential decay over the entire range of $\Delta E$. This is illustrated in Figure~\ref{fig:sigma_vs_dE}. The exponential decay is displayed for large $\Delta E$ on both excitation ($\Delta E < 0$) and quenching ($\Delta E > 0$) wings. In this work, our focus is to exploit this exponential decay of the state-to-state cross sections with $\Delta E$ and use that for our advantage as selection criteria for preparing the training data set for the NNs. A recent study by Joy et al. found a similar trend for H$_{2}$O + H$_{2}$ collisions~\cite{joy2024mixed}. The authors of Ref~\cite{joy2024mixed} fitted their data analytically to compute the coefficients using exponential functions.

\begin{figure}[h]
\centering
  \includegraphics[width=0.5\textwidth, keepaspectratio,]{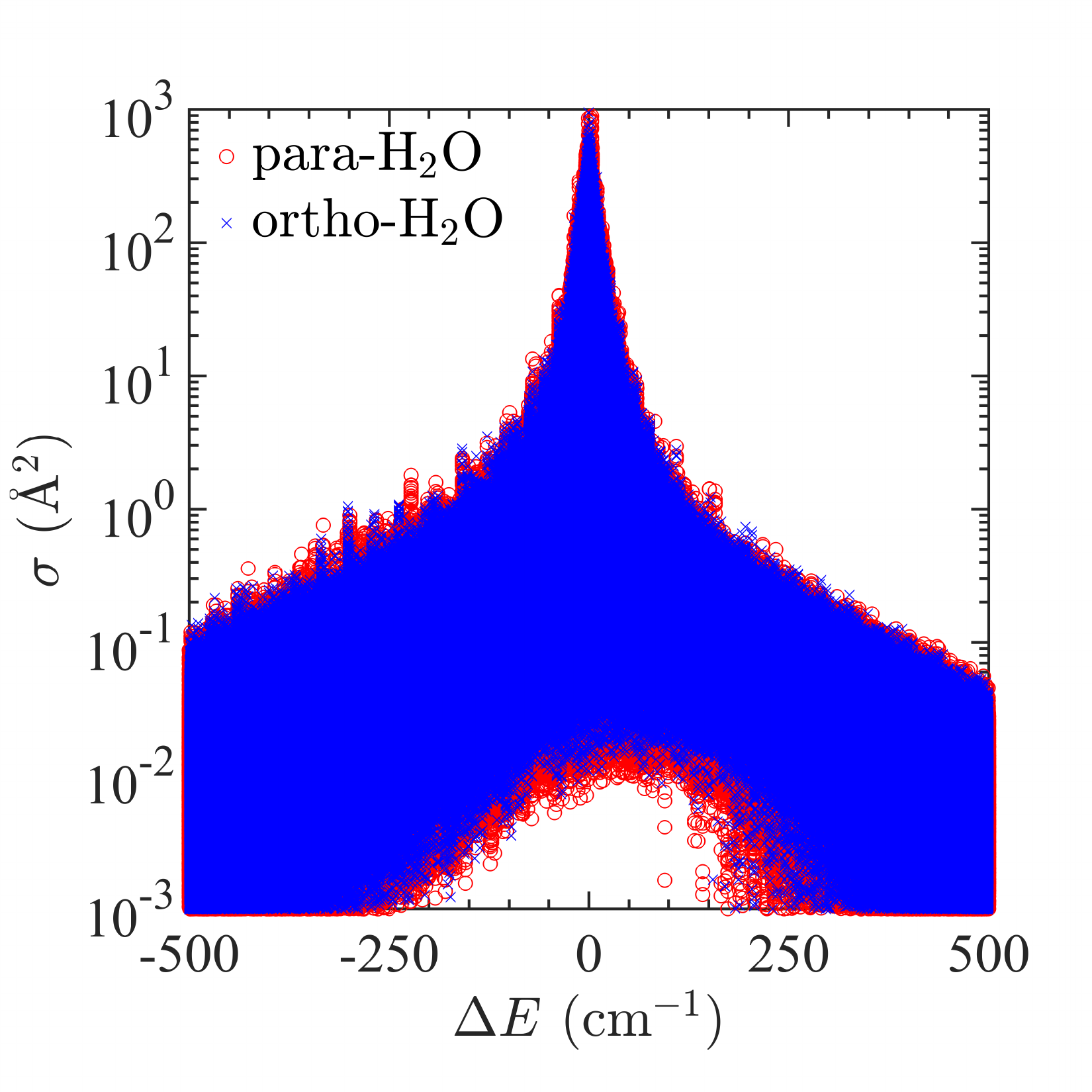}
  \caption{State-to-state cross sections for rotational transitions in H$_{2}$O + H$_{2}$O collisions as functions of the energy difference between initial and final rotational levels ($\Delta E$). Results for para-H$_{2}$O and ortho-H$_{2}$O  targets  are shown by open circles (red) and crosses (blue), respectively.}
  \label{fgr:sigma_vs_dE}
\end{figure}
		
\begin{figure}
	\centering
	\includegraphics[width=0.5\textwidth, keepaspectratio,]{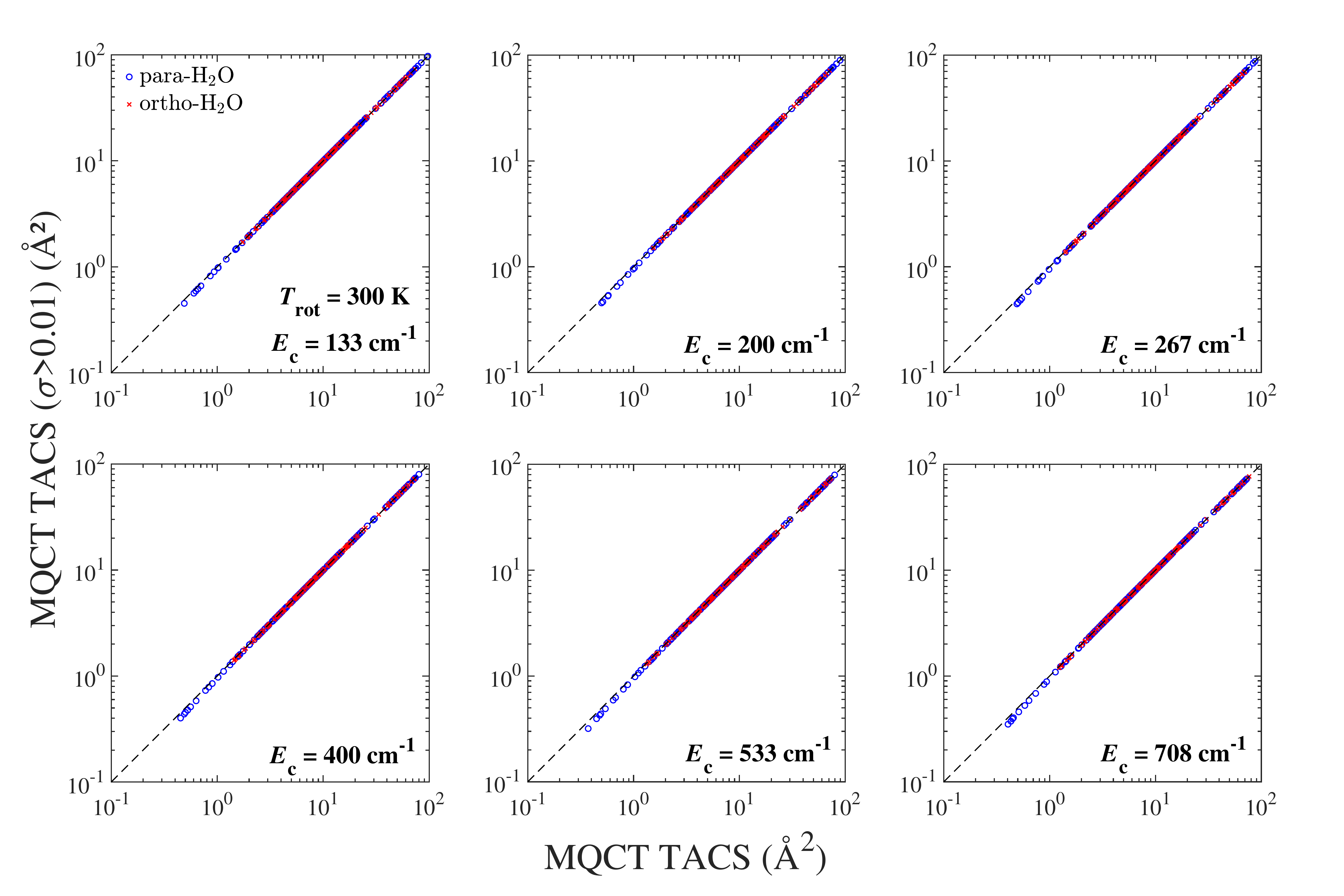}
	\caption{Comparison of TACSs evaluated with all individual state-to-state cross sections  from MQCT calculations with those computed by eliminating $\sigma<0.01~\mathrm{\AA}^{2}$. The dashed black line is the perfect agreement, while blue circles and red crosses correspond to the TACS for para and ortho-H$_{2}$ targets, respectively.}
	\label{fig:TACS_sig_threshold}
\end{figure}

As a result of the exponential decay, the cross sections vary by several orders of magnitude as the energy difference $\Delta E$ increases. Therefore, we start by testing whether the very small cross sections are necessary when the individual cross sections are converted to the TACS. The small-magnitude cross sections are expected to add unnecessary noise to the data, leading to increased complexity of the NNs. Figure~\ref{fig:TACS_sig_threshold} displays a comparison between TACS for state-to-state transitions with all individual cross sections included  compared to  the case in which cross sections smaller than $0.01~\mathrm{\AA}^{2}$ were omitted. The TACS including all individual cross sections are plotted along the horizontal axis, while the TACS computed without $\sigma<0.01~\mathrm{\AA}^{2}$ are plotted along the vertical axis. The black dashed line is the perfect agreement while blue circles and red crosses represent the TACS for para and ortho-H$_{2}$O targets, respectively. It is clear that omitting $\sigma<0.01~\mathrm{\AA}^{2}$ in eq. (\ref{TACS_into_rate}), has no appreciable effect on the accuracy of the TACSs. Therefore, we eliminate individual state-to-state cross sections with magnitude $<0.01~\mathrm{\AA}^{2}$ to build our machine learning models.
		
For training and validation, we take slices from different ranges of $\Delta E$ to make a subset of the entire data as follows,
\begin{equation}
	\begin{aligned}
        \rm{Data}&_{\rm{train-validation}}= \\
	    & \{\sigma_{n_{1}^{}n_{2}^{}\to n_{1}'n_{2}'}(E_{c})\vert \\
	    & (0 \le |\Delta E_{n_{1}^{}n_{2}^{}\to n_{1}'n_{2}'}| \le 10~\mathrm{cm}^{-1})~\land \\
	    & (45 \le |\Delta E_{n_{1}^{}n_{2}^{}\to n_{1}'n_{2}'}| \le 50~\mathrm{cm}^{-1})~\land \\
		& (95 \le |\Delta E_{n_{1}^{}n_{2}^{}\to n_{1}'n_{2}'}| \le 100~\mathrm{cm}^{-1})~\land \\
		& (145 \le |\Delta E_{n_{1}^{}n_{2}^{}\to n_{1}'n_{2}'}| \le 150~\mathrm{cm}^{-1})~\land \\
		& (195 \le |\Delta E_{n_{1}^{}n_{2}^{}\to n_{1}'n_{2}'}| \le 200~\mathrm{cm}^{-1})~\land \\
		& (245 \le |\Delta E_{n_{1}^{}n_{2}^{}\to n_{1}'n_{2}'}| \le 250~\mathrm{cm}^{-1})~\land \\
		& (295 \le |\Delta E_{n_{1}^{}n_{2}^{}\to n_{1}'n_{2}'}| \le 300~\mathrm{cm}^{-1})~\land \\
		& . \\
		& . \\
		& . \\
		& \}.
	\end{aligned}
	\label{train_val_data}
\end{equation}
If the smallest and largest value of $|\Delta E|$ lies outside of a given slice, then they are added to this subset. About 80\% of this subset is used for training and the remaining 20\% for validation through random sampling. The remaining data outside of this aforementioned subset are used for testing.
		 
\begin{figure}
	\centering
	\includegraphics[width=0.5\textwidth, keepaspectratio,]{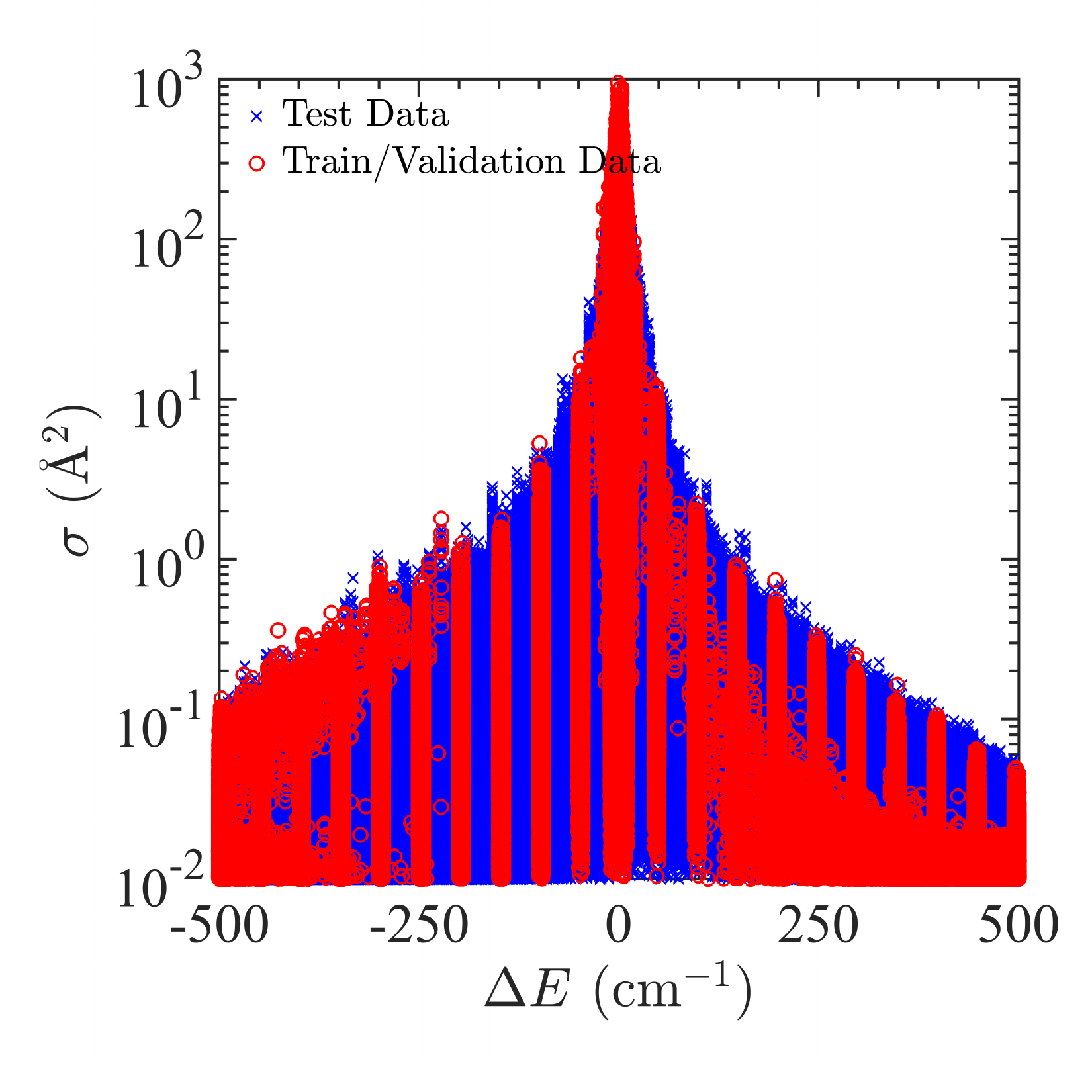}
	\caption{A visual representation of the data used for training and validation as well as testing. The red circles denote the data for training and validation and the blue crosses denote the data for testing as given by eq. (\ref{train_val_data}).}
	\label{fig:sigma_vs_dE_step50}
\end{figure}

This dataset for training and validation as well as the test dataset are displayed in Figure~\ref{fig:sigma_vs_dE_step50} as functions of $\Delta E$. The red circles represent this subset of data for training and validation, while the blue crosses represent the remaining data that are used for testing. As Figure~\ref{fig:sigma_vs_dE_step50} illustrates, for every $50$ cm$^{-1}$ of $\Delta E$, there are red patches that denote the data specified in eq. (\ref{train_val_data}). However, some additional data (red circles) are also visible in the figure which do not fall into this category of $50$ cm$^{-1}$ of $\Delta E$. These data points correspond to the smallest and largest values of $\Delta E$ for each of the initial rotational states as explained previously.
		 
In general, we aim to design the training and validation dataset that consists of cross section values for which the energy difference of the transition lies within the entire range so that there are no predictions to be made outside the range of the training data. All the cross sections for the remaining transitions are used as the test data set. In this work, multiple slices are made through the whole data set for ML models to optimize the computational efficiency and accuracy of neural networks (NNs), as discussed later.
		
Most importantly, we found that the excitation wing ($\Delta E < 0$) display subtle differences in behavior compared to the quenching ($\Delta E > 0$) wing,  so separate models were built for these two regimes. In this context, one should note that each of the collision energies as well as para and ortho-H$_{2}$O symmetry were treated separately to build a separate NN. We built 4 models using NN for a given set of training and validation data corresponding to combinations of the excitation and quenching transitions for both para and ortho symmetry for each of the six collision energies, which leads to a total of 24 models.
		
Each of our dataset has thirteen features as input parameters for the NNs: rotational quantum numbers of initial and final states of the first water molecule ($j_{1}k_{a_{1}}k_{c_{1}}, j'_{1}k'_{a_{1}}k'_{c_{1}}$),  the second water molecule ($j_{2}k_{a_{2}}k_{c_{2}}, j'_{2}k'_{a_{2}}k'_{c_{2}}$), and the energy difference between the initial and final states of the molecular system, $\Delta E$. The NNs are designed to interpolate over all these input features. Since the input features are comprised of different data types (integers for rotational quantum numbers of the initial and final states while float for the energy difference) with a large variation in the magnitude of the input data, they needed to be scaled for the NNs to work optimally. This is done by using ``StandardScaler" function from the ``sklearn" package to have zero-mean and unit-variance as $\tilde{x} = \frac{x-u}{s}$, where $u$ and $s$ are the mean and the standard deviation of the features, respectively. This standardization of the input data is done so that none of the features get higher weights just because of their magnitude being larger than the values of other features. Note that this scaling is applied only to the input features as listed previously, and not the cross section, $i.e.$, output. The same transformation is applied uniformly to all three data sets: training, validation and testing.
				
In our data analysis, we found that the dependent feature, $i.e.,$ cross sections for individual state-to-state rotational transitions, vary by several orders of magnitude. We found that the NNs do not perform well for data that vary over several orders of magnitude. To resolve this issue, we used logarithm of the cross sections in our ML modeling [$y=\log_{10}(\sigma)$] as target. The predictions are then converted back to cross sections as $\sigma=10^{y}$.
		
\subsubsection{NN details}
\begin{figure}
	\centering
	\includegraphics[width=0.5\textwidth, keepaspectratio,]{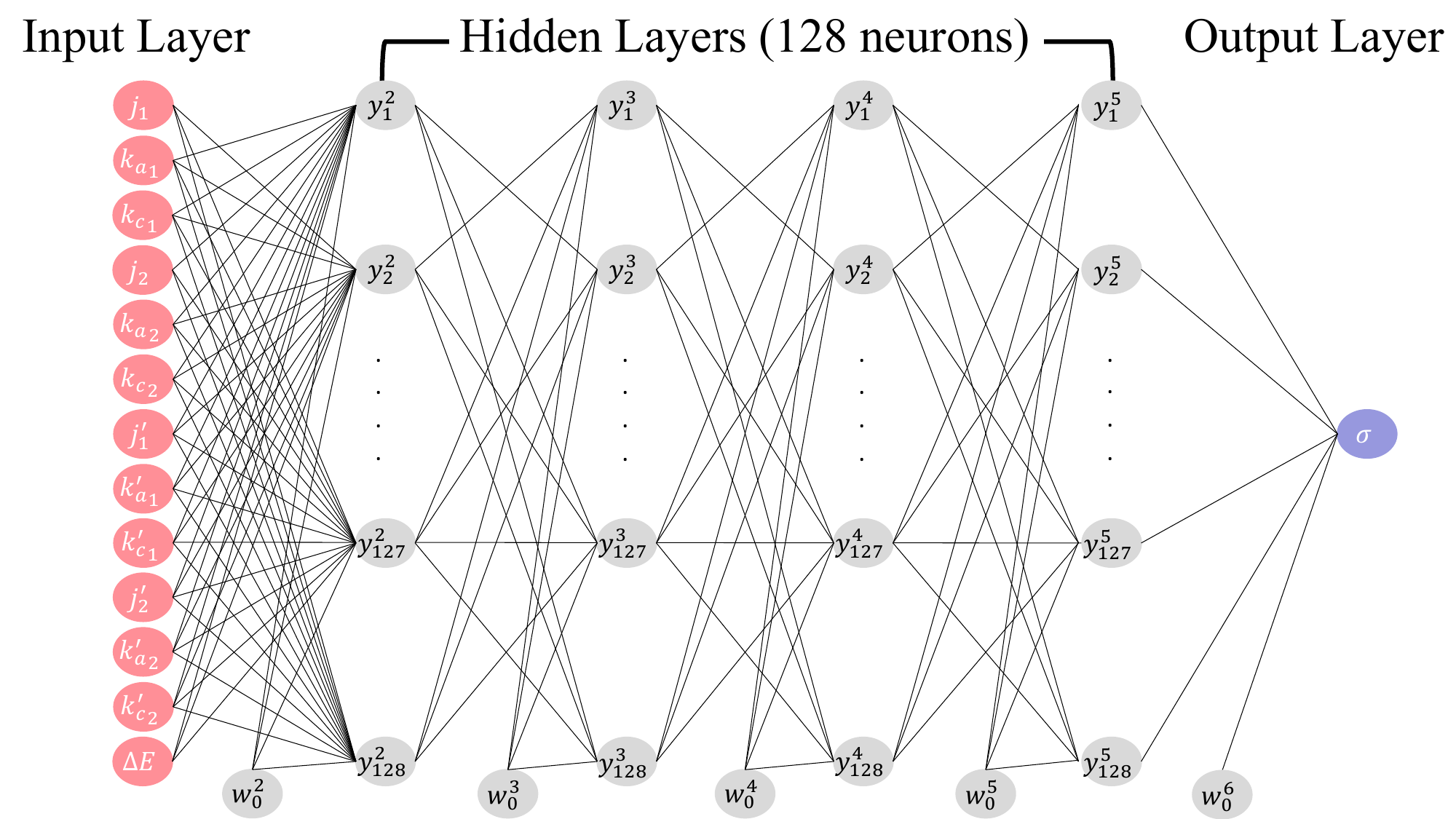}
	\caption{Architecture of the ML models comprised of four hidden layers with each having 128 neurons, thirteen features for the input layer, and one output neuron for the logarithm of the cross sections.}
	\label{fig:MLP_fig}
\end{figure}

The NNs all have the same architecture, each characterized by one input layer with thirteen neurons corresponding to the specific features of our dataset. The optimal number of hidden layers following the input layer was determined through exploratory search to optimize the model performance. The root mean squared error or RMSE for the test data corresponding to two, three and four hidden layers, each with 128 neurons, were respectively $0.81$, $0.74$ and $0.73~\mathrm{\AA}^{2}$. Therefore, we decided to build our models with four hidden layers, each with 128 neurons. There is one output layer with a single neuron corresponding to the logarithm of cross section. A schematic diagram of our NNs is shown in Figure~\ref{fig:MLP_fig}.
		
The rectified linear unit (ReLU) $[f(x)=\mathrm{max}(0, x)]$ is used throughout all the hidden layers as activation function. We used the Adam optimizer for training our NNs with a learning rate of 0.0001~\cite{kingma2014adam}. The details of the NNs including the number of parameters for each layer are provided in Table \ref{table:MLP_model}.
		
\begin{table}
	\caption{\label{table:MLP_model} Summary of the architecture and technical details of the NNs.}
	\centering
	\resizebox{\linewidth}{!}{%
	\begin{tabular}{|l|c|c|c|}
		\hline
		Layer name & No. of neurons & No. of parameters & Activation \\ \hline
		Input & 13 & 13 &  \\ \hline
		Hidden layer 1: Dense & 128 & 1792 & ReLU \\ \hline
		Hidden layer 2: Dense & 128 & 16512 & ReLU \\ \hline
		Hidden layer 3: Dense & 128 & 16512 & ReLU \\ \hline
		Hidden layer 4: Dense & 128 & 16512 & ReLU \\ \hline
		Output & 1 & 129 & \\ \hline
	\end{tabular}%
	}
\end{table}
		
Our models were built using TensorFlow with a batch size of 32 and a maximum of 300 epochs~\cite{abadi2015tensorflow}. An early stopping mechanism with a patience of 30 using the validation dataset was adopted to prevent the models from overfitting. Additionally, we used ridge regularization ($i.e.$, L2 regularization) with a penalty of 0.01 to the weights of the kernels for each hidden layer. This discourages large weights and reduces the complexity of our NNs. A summary of the NN hyperparameters is provided in Table \ref{table:parameter_summary}.
		
\begin{table}
	\caption{\label{table:parameter_summary} A summary of the hyper-parameters and their values used for training the ML models}
	\centering
	\resizebox{\linewidth}{!}{%
	\begin{tabular}{|l|c|c|}
		\hline
		Hyperparameter & Type/Value & Additional details \\ \hline
		Batch size & 32 & \\ \hline
		Optimizer & Adam~\cite{kingma2014adam} & Learning rate = 0.0001\\ \hline
		Kernel regularization & Ridge (L2) & Regularization weight = 0.01\\ \hline
		Maximum no. of epochs & 300 & \\ \hline
		Early stopping & Implemented & Patience = 30 \\ \hline
	\end{tabular}%
	}
\end{table}
		
The mean-squared error function, ${\rm{MSE}} = \frac{1}{n}\sum_{i=1}^{n}(y_{\rm{MQCT}}^i - y_{\rm{predicted}}^i)^{2}$, was used as the loss function to quantitatively analyze the performance of our ML models applied to both the training data set as well as the validation data set. Here, $n$ is the total number of samples used, the variables $y_{\rm{MQCT}}^i$ and $y_{\rm{predicted}}^i$ represent the logarithm of the actual cross sections from MQCT calculations and the cross sections predicted by the NNs, respectively. We also monitored the RMSE $= \sqrt{\rm{MSE}}$ for the interpretation of the predicted data.
		
\section{Results}
\label{sec:results}
\subsection{Optimization of dataset for training, validation \& testing}

First, we analyzed the whole dataset of individual state-to-state cross sections for rotational transitions in collision of two H$_{2}$O molecules. This dataset included all combinations of ortho- and para-H$_{2}$O, $i.e.$, ortho-ortho, ortho-para, para-ortho, para-para combinations considering both target and quencher H$_{2}$O molecules and for all six collision energies. Figure~\ref{fig:sigma_vs_dE} displays this  cross section data as a function of the energy difference ($\Delta E$) between the initial and final rotational states. Cross sections for para-H$_{2}$O as target molecule are shown by red open circles, while blue crosses represent the same for ortho-H$_{2}$O as the target. They are characterized by  a single exponential decay for	$|\Delta E| \le 50$ cm$^{-1}$ followed by a second exponential decay as  $|\Delta E|$ increases.
		
To adequately capture the exponential decay of the cross sections with $\Delta E$, we adopted different slices of the entire data set for training and validation and built several NNs as part of the optimization process. The accuracy of the models was determined at two different levels. First, all the individual state-to-state transitions were tested against the whole test dataset. Second, thermally averaged cross sections were computed using the NN predictions and compared against the actual MQCT TACS.
		
\textbf{Dataset 1.} We started to build our NNs using the data shown in Figure~\ref{fig:sigma_vs_dE_step50} and specified by eq. (\ref{train_val_data}). The predicted cross sections for state-to-state transitions were compared against the actual MQCT cross sections from the test data for both para and ortho-H$_{2}$O targets as shown in Figure S1 and Figure S2, respectively, of the Supplementary Material. The resulting TACSs using these predicted data are compared against the actual MQCT TACSs, in Figure S3 of the Supplementary Material. Note that the computation of TACSs requires both the NN predictions and the data used for training and validation. The agreement at the level of individual cross sections is found to be reasonable for both para and ortho-H$_{2}$O targets, while the agreement at the level of TACS is found to be excellent.
		
\textbf{Dataset 2.} We explored if we can reduce the training dataset, while preserving model accuracy, to improve efficiency. Therefore, instead of composing the subset data for training and validation at every $\Delta E=50$ cm$^{-1}$, we built a subset with the step of 100 cm$^{-1}$ in $\Delta E$:
\begin{equation}
	\begin{aligned}
		\rm{Data}&_{\rm{train-validation}}=&\\
		& \{\sigma_{n_{1}^{}n_{2}^{}\to n_{1}'n_{2}'}(E_{c})\vert\\
		& (0 \le |\Delta E_{n_{1}^{}n_{2}^{}\to n_{1}'n_{2}'}| \le 10~\mathrm{cm}^{-1})~\land \\
		& (95 \le |\Delta E_{n_{1}^{}n_{2}^{}\to n_{1}'n_{2}'}| \le 100~\mathrm{cm}^{-1})~\land \\
		& (195 \le |\Delta E_{n_{1}^{}n_{2}^{}\to n_{1}'n_{2}'}| \le 200~\mathrm{cm}^{-1})~\land \\
		& (295 \le |\Delta E_{n_{1}^{}n_{2}^{}\to n_{1}'n_{2}'}| \le 300~\mathrm{cm}^{-1})~\land \\
		& . \\
		& . \\
		& \}
	\end{aligned}
	\label{train_val_data_100}
\end{equation}
		
The sets of training and test data from Dataset 2 are displayed in Figure S4. The comparison with the actual data for individual state-to-state cross sections became slightly worse for both para and ortho-H$_{2}$O targets as shown in Figure S5 and Figure S6, respectively. The predicted TACS also displays larger discrepancies with the MQCT TACS as shown in Figure S7 of the Supplementary Material.
		
\textbf{Dataset 3.} The training data needs to reflect Dataset 1 due to its double exponential feature. The models need to capture the fact that the slope changes depending on the range of $\Delta E$ and the training data should contain that feature. As explained earlier, the slope of the exponential decay changes rapidly near $\Delta E=50$ cm$^{-1}$, and so this data should be part of the training set as shown in eq. (\ref{train_val_data_mixed_300}) below. Moreover, we explored reducing the range of the maximum magnitude of $\Delta E$ to check if we really need to sample from the entire data. We systematically reduced the value of $\Delta E$ and found that eliminating data above $\vert\Delta E\vert\ge300$ cm$^{-1}$ has  minimal effect on the overall TACS. We thus arrived at the following dataset:
		
\begin{equation}
	\begin{aligned}
		\rm{Data}&_{\rm{train-validation}}=\\
		& \{\sigma_{n_{1}^{}n_{2}^{}\to n_{1}'n_{2}'}(E_{c})\vert\\
		& (0 \le |\Delta E_{n_{1}^{}n_{2}^{}\to n_{1}'n_{2}'}| \le 10~\mathrm{cm}^{-1})~\land \\
		& (45 \le |\Delta E_{n_{1}^{}n_{2}^{}\to n_{1}'n_{2}'}| \le 50~\mathrm{cm}^{-1})~\land \\
		& (95 \le |\Delta E_{n_{1}^{}n_{2}^{}\to n_{1}'n_{2}'}| \le 100~\mathrm{cm}^{-1})~\land \\
		& (145 \le |\Delta E_{n_{1}^{}n_{2}^{}\to n_{1}'n_{2}'}| \le 150~\mathrm{cm}^{-1})~\land \\
		& (195 \le |\Delta E_{n_{1}^{}n_{2}^{}\to n_{1}'n_{2}'}| \le 200~\mathrm{cm}^{-1})~\land \\
		& (245 \le |\Delta E_{n_{1}^{}n_{2}^{}\to n_{1}'n_{2}'}| \le 250~\mathrm{cm}^{-1})~\land \\
		& (295 \le |\Delta E_{n_{1}^{}n_{2}^{}\to n_{1}'n_{2}'}| \le 300~\mathrm{cm}^{-1}) \}
	\end{aligned}
	\label{train_val_data_mixed_300}
\end{equation}
		
\begin{figure}
	\centering
	\includegraphics[width=0.5\textwidth, keepaspectratio,]{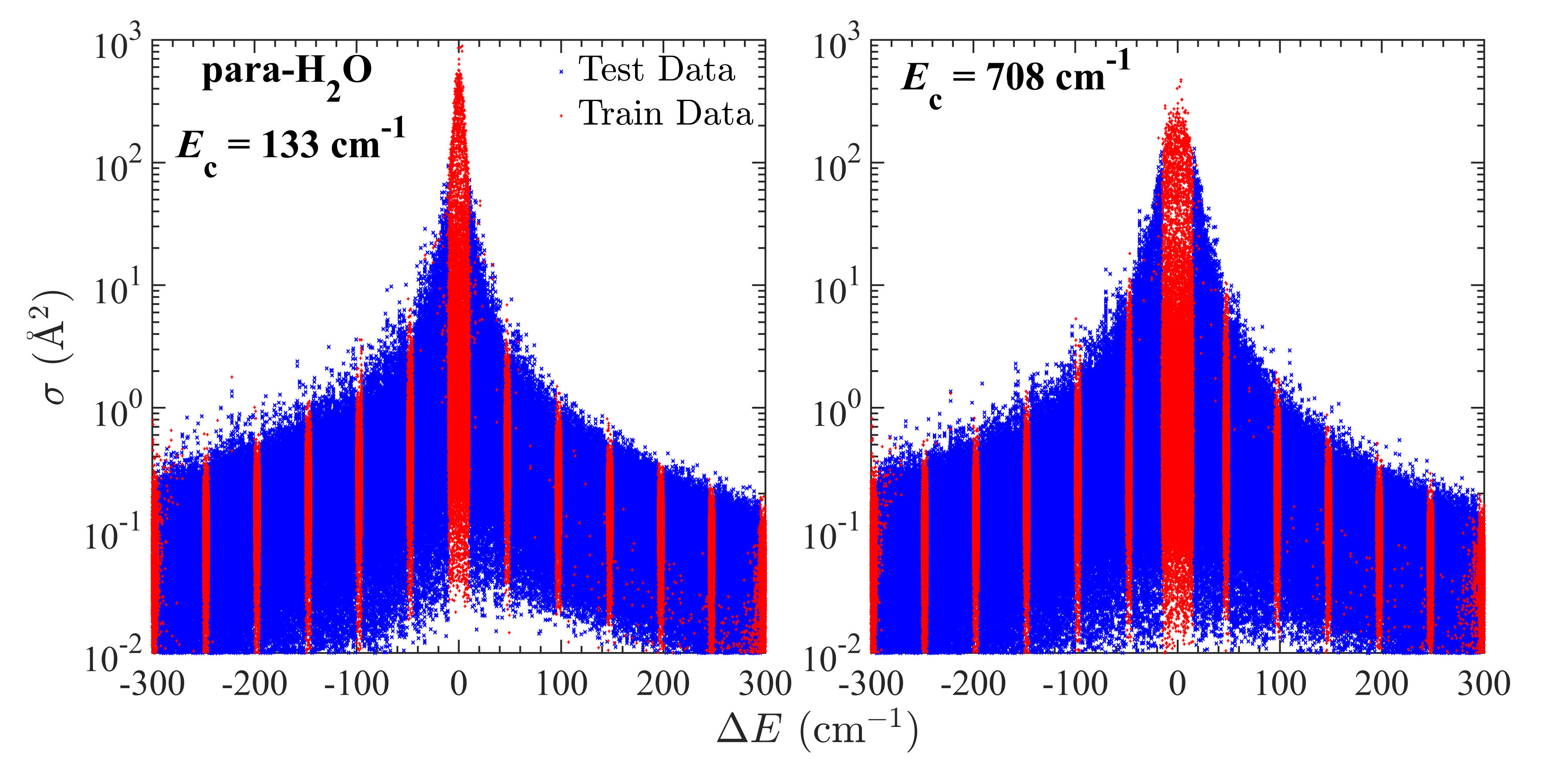}
	\caption{The training and test data for all state-to-state transitions at the highest and lowest collision energies, based on Dataset 3, are presented as a function of $\Delta E$ for the para-H$_{2}$O molecule.  The dataset for other collision energies and for ortho-H$_{2}$O is very similar.}
	\label{fig:sigma_vs_dE300_para}
\end{figure}

This is the training/validation set that we label as the best performing and the most optimized for reliable predictions. We adopt this data structure shown by eq. (\ref{train_val_data_mixed_300}) for collision energies 133, 200, and 267 cm$^{-1}$. Because the density of individual cross sections near the elastic peak decreases rapidly with collision energy, for $E_{c}=400$ cm$^{-1}$ and higher, we replace  $\Delta E<10$ cm$^{-1}$ by $\Delta E<15$ cm$^{-1}$ in eq.(\ref{train_val_data_mixed_300}) to have adequate sampling near $\Delta E=0$ cm$^{-1}$.	The resulting training and test data corresponding to individual state-to-state rotational transitions are displayed in Figure~\ref{fig:sigma_vs_dE300_para} for para-H$_{2}$O molecule at the highest and lowest collision energies. For other collision energies and ortho-H$_{2}$O molecules, the dataset are very similar.

A summary of the sizes of the training, validation and test datasets from eq. (\ref{train_val_data_mixed_300}) is given in Table~\ref{table:dataset_summary} that includes both exchange symmetry of para and ortho-H$_{2}$O molecule and both excitation and quenching transitions. The size of the training data is about $\sim$10\% of the entire dataset after removing the small-magnitude cross sections and limiting the range of energy gap to $\Delta E\le300$ cm$^{-1}$. The size of the training, validation and test datasets remains almost the same for the lower three collision energies for both para and ortho symmetries and for both excitation and quenching transitions. The same applies to the higher three collision energies, but the size of the training, validation and test datasets are slightly different as explained before.
		
\begin{table}
	\caption{\label{table:dataset_summary} A summary of the sizes of training, validation and test datasets for both para- and ortho-  H$_{2}$O molecules and for both excitation ($\Delta E < 0$) and quenching ($\Delta E > 0$) transitions are listed. The corresponding relative RMSE is also listed in the last column.}
	\centering
	\resizebox{\linewidth}{!}{
		\begin{tabular}{|c|c|c|c|c|c|c|}
			\hline
			$E_{\mathrm{c}}$ & H$_{2}$O & $\Delta E$ & Training & Validation & Test & Relative\\ 
			(cm$^{-1}$) & Symmetry & sign & Data Size & Data Size & Data Size & RMSE \\ \hline \hline
			133 & para & (+ve) & 34422 & 7772 & 273223 & 0.430 \\ \hline
			200 & para & (+ve) & 34424 & 7772 & 273232 & 0.379 \\ \hline
			267 & para & (+ve) & 34445 & 7778 & 273374 & 0.421 \\ \hline
			400 & para & (+ve) & 39165 & 8957 & 267713 & 0.402 \\ \hline
			533 & para & (+ve) & 39162 & 8957 & 267783 & 0.404 \\ \hline
			708 & para & (+ve) & 39199 & 8966 & 267967 & 0.397 \\ \hline \hline
			133 & ortho & (+ve) & 33161 & 7494 & 259189 & 0.429 \\ \hline
			200 & ortho & (+ve) & 33164 & 7495 & 259241 & 0.384 \\ \hline
			267 & ortho & (+ve) & 33186 & 7501 & 259329 & 0.389 \\ \hline
			400 & ortho & (+ve) & 37526 & 8586 & 254107 & 0.387 \\ \hline
			533 & ortho & (+ve) & 37541 & 8590 & 254221 & 0.380 \\ \hline
			708 & ortho & (+ve) & 37552 & 8592 & 254307 & 0.367 \\ \hline \hline
			133 & para & ($-$ve) & 36196 & 8213 & 291562 & 0.438 \\ \hline
			200 & para & ($-$ve) & 36137 & 8199 & 291201 & 0.442 \\ \hline
			267 & para & ($-$ve) & 36161 & 8205 & 291181 & 0.462 \\ \hline
			400 & para & ($-$ve) & 40837 & 9374 & 285384 & 0.414 \\ \hline
			533 & para & ($-$ve) & 40864 & 9380 & 285566 & 0.431 \\ \hline
			708 & para & ($-$ve) & 40891 & 9387 & 285943 & 0.404 \\ \hline \hline
			133 & ortho & ($-$ve) & 35544 & 8089 & 280565 & 0.444 \\ \hline
			200 & ortho & ($-$ve) & 35525 & 8084 & 280295 & 0.430 \\ \hline
			267 & ortho & ($-$ve) & 35528 & 8085 & 280334 & 0.416 \\ \hline
			400 & ortho & ($-$ve) & 39844 & 9163 & 275045 & 0.424 \\ \hline
			533 & ortho & ($-$ve) & 39859 & 9167 & 275191 & 0.409 \\ \hline
			708 & ortho & ($-$ve) & 39880 & 9172 & 275370 & 0.381 \\ \hline
		\end{tabular}
	}
\end{table}
		
\begin{figure}
	\centering
	\includegraphics[width=0.5\textwidth, keepaspectratio,]{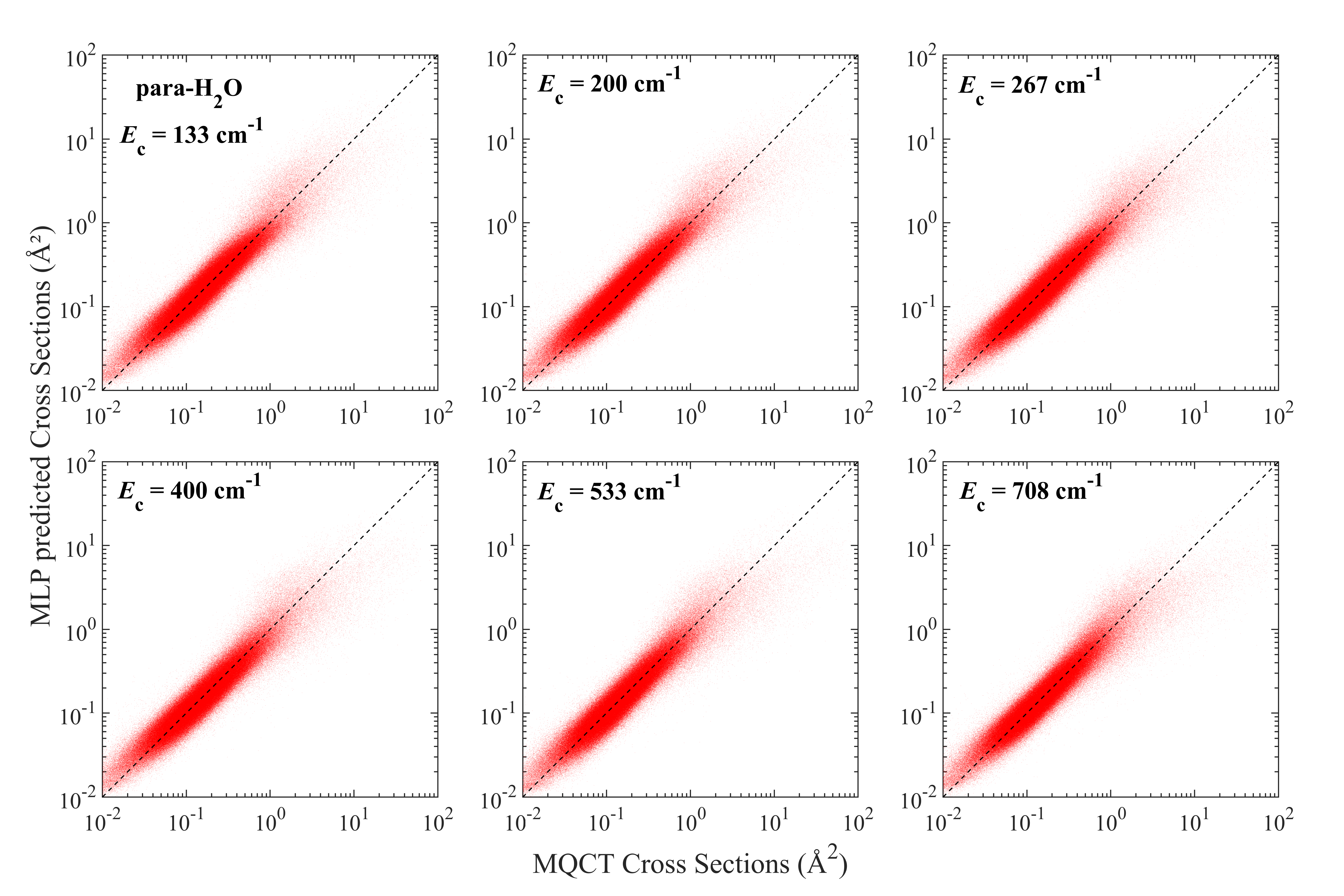}
	\caption{A comparison of ML predicted state-to-state cross sections against the actual MQCT results for para-H$_{2}$O molecule.}
	\label{fig:sigma_para_dE300}
\end{figure}

\begin{figure}
	\centering
	\includegraphics[width=0.5\textwidth, keepaspectratio,]{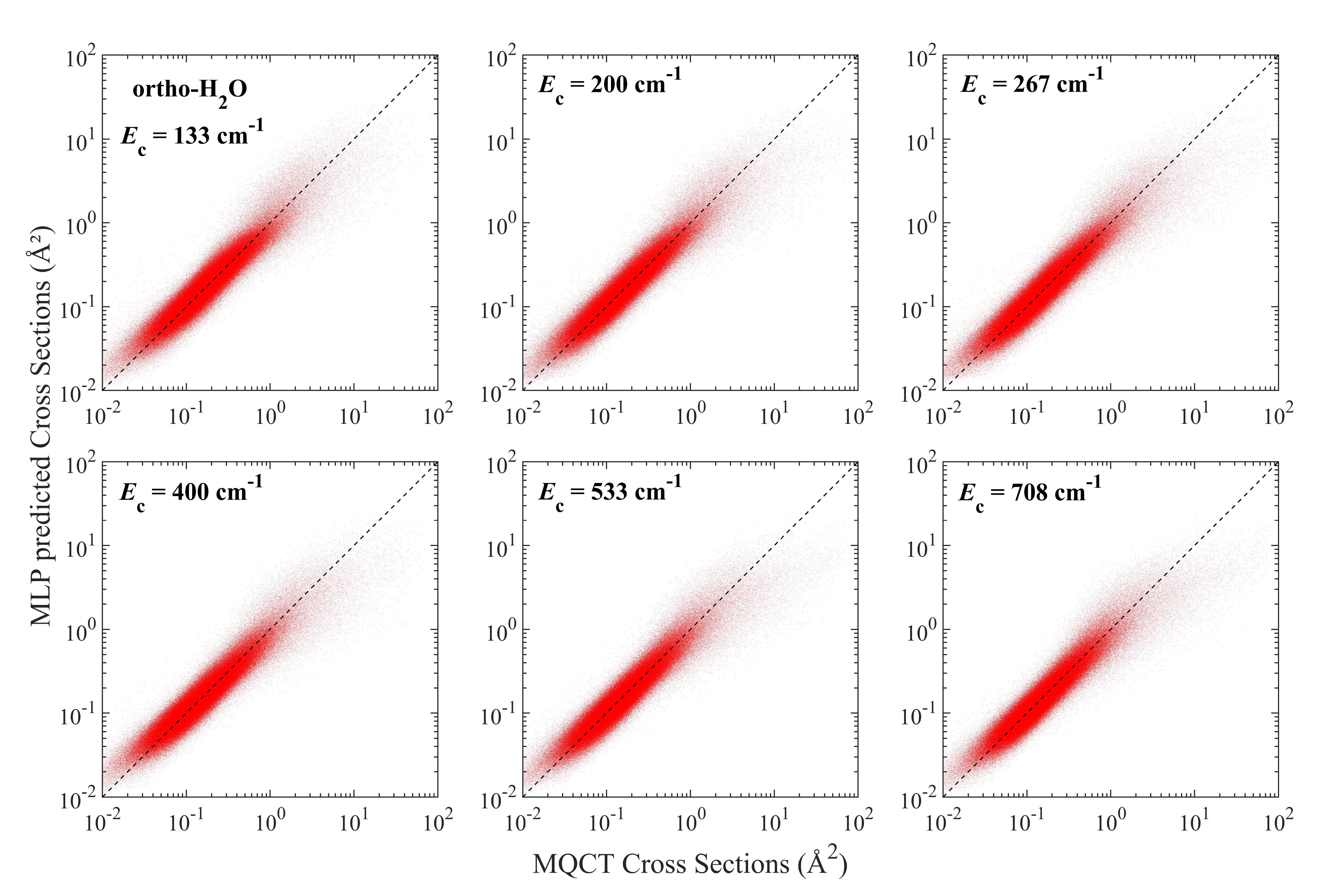}
	\caption{Same as Figure~\ref{fig:sigma_para_dE300}, but for ortho-H$_{2}$O molecule.}
	\label{fig:sigma_ortho_dE300}
\end{figure}

The predicted cross sections for the individual state-to-state rotational transitions of both para and ortho-H$_{2}$O molecules are shown in Figure~\ref{fig:sigma_para_dE300} and \ref{fig:sigma_ortho_dE300}, respectively. The horizontal axis represents the MQCT cross sections while the NN predictions are plotted along the vertical axis. The black dashed line would be the perfect agreement while the red dots represent the comparison. It is seen that  the predicted cross sections agree reasonably well with the MQCT cross sections and are within the range of acceptable accuracy for both para and ortho symmetry of the H$_{2}$O molecule. While the smaller cross sections (corresponding to larger $\Delta E$) exhibit higher discrepancies (relative to their magnitudes), their overall contribution to TACS is less significant.
		
To further quantify the errors in the NN predictions for each collision energy, specific symmetry of the H$_{2}$O molecule (para or ortho), and excitation/quenching transitions ($\Delta E < 0$ or $\Delta E > 0$), we report in the last column of Table~\ref{table:dataset_summary} the relative RMSE or RRMSE, defined as:
		
\begin{equation}
	\mathrm{RRMSE} = \sqrt{\frac{1}{N}\sum_{i=1}^{N}\left(\frac{\sigma^{i}_{\mathrm{MQCT}}-\sigma^{i}_{\mathrm{predicted}}}{\sigma^{i}_{\mathrm{MQCT}}} \right)^2},
	\label{rrmse}
\end{equation}
where $\sigma^{i}_{\mathrm{MQCT}}$ refers to the MQCT cross sections not used for training and validation. The RRMSE values range from  $\sim$37\% to $\sim$46\% with an average value of $\sim$41\%.
		
We have also examined whether similar level of accuracy can be reached with a fewer number of hidden layers and number of neurons in each layer to reduce the complexity of the NNs since the training dataset in this case is relatively small compared to Dataset 1. We found that reducing the number of hidden layers does not significantly make the prediction worse with respect to the RMSE values nor meaningfully improve the computational efficiency. Therefore, we retained four hidden layers with 128 neurons each.
		
\begin{figure}
	\centering
	\includegraphics[width=0.5\textwidth, keepaspectratio,]{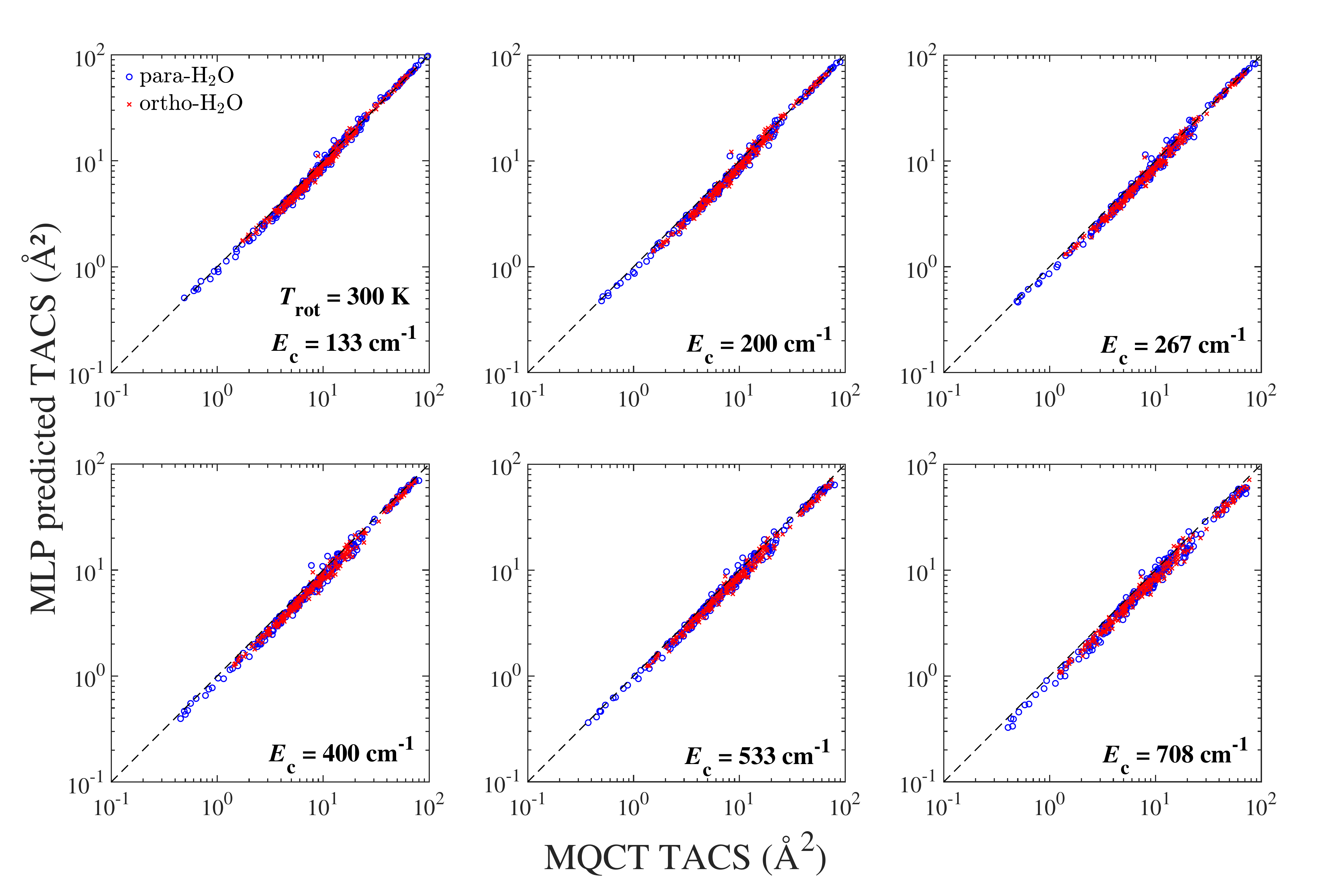}
	\caption{A comparison of thermally averaged cross sections (TACSs) computed using actual MQCT state-to-state transitions cross sections and our NN predicted cross sections is displayed here for six different collision energies as labeled in each frame. Blue empty circles, and red crosses correspond to transitions between target para-H$_{2}$O and ortho-H$_{2}$O molecules.}
	\label{fig:T_300_dE300}
\end{figure}
		
Finally, we composed the TACS following eq. (\ref{TACS_eq}) using the predicted individual state-to-state transitions for the test data combined with original training and validation data and compared against the actual MQCT TACS. Figure~\ref{fig:T_300_dE300} displays this comparison. The blue circles and red crosses correspond to the para and ortho symmetry of the H$_{2}$O molecule, respectively. These TACSs are computed for a rotational temperature $T_{\rm{rot}}=300$ K. The agreement is excellent between the actual MQCT TACS and that NN predictions. The agreement does not improve significantly when more data is included by extending the range of the $\Delta E$ as shown in Figure S3 of the Supplementary Material using Dataset 1.
		
\begin{figure}
	\centering
	\includegraphics[width=0.5\textwidth, keepaspectratio,]{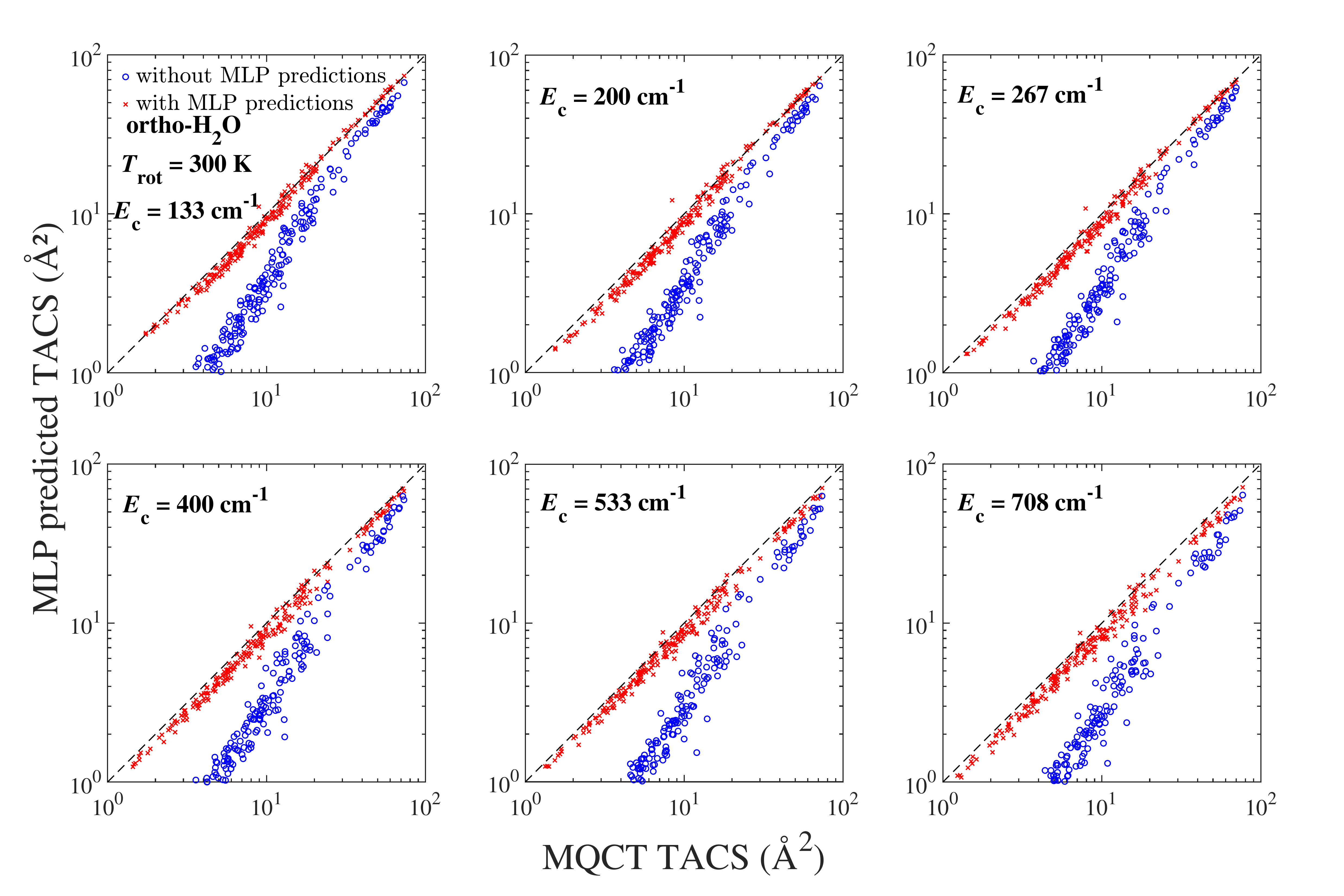}
	\caption{Contribution of ML predicted cross sections toward the TACS  by excluding and including the predicted test data for ortho-H$_{2}$O molecules by blue circles and red crosses, respectively. This red crosses are same as in Fig.~\ref{fig:T_300_dE300}.}
	\label{fig:T_300_dE300_ortho}
\end{figure}
		
It should be noted that the original MQCT state-to-state cross sections were divided into a training and validation set (about 10\%) and a test set (about 90\%). In the comparison provided in Fig. \ref{fig:T_300_dE300} the test set was replaced by the ML produced data. To demonstrate the important contributions of the NN predicted data toward the overall TACSs, we computed the TACS by using only the training and validation data. The resulting TACS are compared against those obtained by including the NN predictions and the actual MQCT TACS in Figure~\ref{fig:T_300_dE300_ortho} for ortho-H$_{2}$O. Similar results  for para-H$_{2}$O are provided in Figure S8 of the Supplementary Material. The TACS using all of the original MQCT cross sections are plotted along the horizontal axis, while the red crosses and blue circles correspond to the TACS computed with and without incorporating the NN predictions into the training and validation data. The dashed black curve would represent the perfect agreement. These red crosses are the same as shown in Fig.~\ref{fig:T_300_dE300}. It can be seen that the blue circles deviate significantly from the perfect agreement represented by the black dashed diagonal line. Thus, the contribution of the NN-predicted cross sections is significant accounting for nearly 90\% of the original MQCT cross sections not used for training and validation. Thus, the approach presented here can reliably be applied to other complex molecular systems with significant savings in computational efforts.
		
The TACSs are one of the main ingredients for astrophysical models and serve as an important input to the numerical codes of radiative transfer modeling, such as RADEX, MOLPOP or LIME. To confirm that the accuracy of the NNs implemented here is significant for these models, we computed the percentage deviation of the TACSs as predicted by our NNs and the original MQCT TACSs. The computed percentage deviation is then averaged over all transitions for target H$_{2}$O molecules over 231 para-H$_{2}$O and 210 ortho-H$_{2}$O transitions. The resulting data, presented in Table~\ref{table:percent_diff_TACS},  illustrate that the agreement is excellent at the level of TACS despite larger deviations at the state-to-state level. Because TACSs are the main quantity that is relevant for  modeling energy transfer in  astrophysical environments, the level of accuracy attained in our models is adequate for astrophysical models. The average percentage deviation is about $\sim$13\% and $\sim$14\%, while the largest error is about $\sim$15\% and $\sim$17\% for para and ortho-H$_{2}$O targets, respectively. This is very encouraging since our goal is to reduce the requirements of the computational resources to explicitly compute the relevant TACSs. This is achieved in our proposed workflow without losing  significant accuracy but at very low computational cost.
		
\begin{table}
	\caption{\label{table:percent_diff_TACS} Average percent deviation between the ML predicted TACS and the MQCT TACS for both ortho- and para- H$_{2}$O targets at different collision energies.}
	\centering
	\makebox[\linewidth]{
		\renewcommand{\arraystretch}{1.1}
		\begin{tabular}{|c|c|c|}
			\hline
			$E_{\mathrm{c}}$ & \multicolumn{2}{c|}{Average \% difference} \\ \cline{2-3}
			(cm$^{-1}$) & para-H$_{2}$O & ortho-H$_{2}$O \\ \hline
			133 & 11.87 & 10.00 \\ \hline
			200 & 13.80 & 13.73 \\ \hline
			267 & 12.50 & 14.18 \\ \hline
			400 & 13.06 & 14.72 \\ \hline
			533 & 13.30 & 14.28 \\ \hline
			708 & 15.00 & 16.99 \\ \hline
		\end{tabular}
	}
\end{table}
		
\begin{figure}
	\centering
	\includegraphics[width=0.5\textwidth, keepaspectratio,]{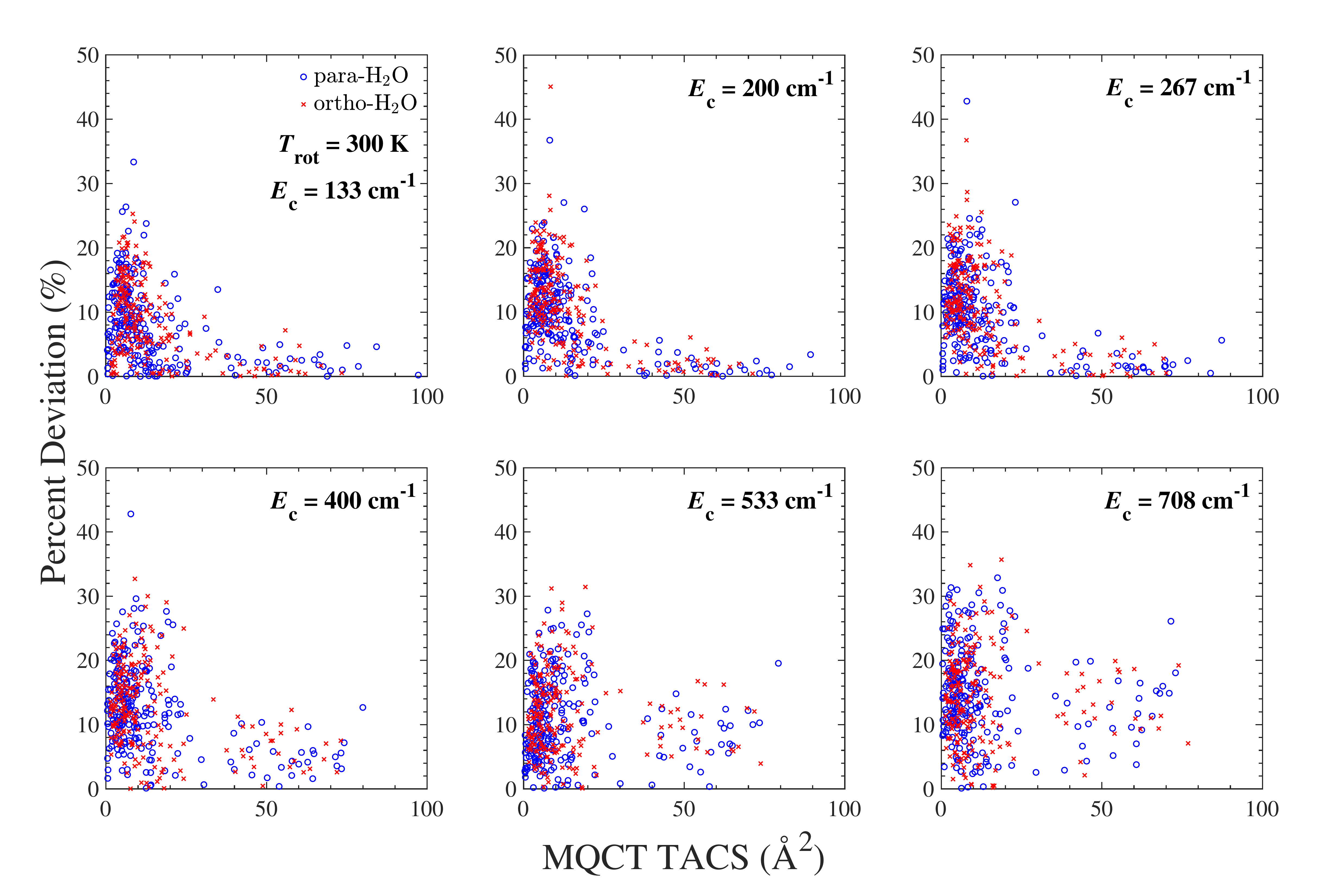}
	\caption{Percentage deviation of the ML predicted TACS from the actual MQCT TACS as a function of the magnitude of the MQCT TACS.}
	\label{fig:T_300_dE300_percent}
\end{figure}
			
To illustrate the accuracy of TACS derived from the NNs, a plot of percent deviation for all 441 transitions of target H$_{2}$O molecule considering both ortho- and para-H$_{2}$O symmetry for all six collision energies is displayed in Figure~\ref{fig:T_300_dE300_percent} as a function of the magnitude of actual MQCT TACS. It is seen that for larger values of TACSs, the percentage deviation remains in the range of $\sim$10-20\% considering all collision energies and both symmetries of target H$_{2}$O molecule. The agreement is better for lower values of collision energies and decreases slightly at higher collision energies. We also notice an increase in the percentage deviation for lower magnitude of the TACSs for both para and ortho-H$_{2}$O transitions, which is expected.
		
\textbf{Dataset 4.}  Because the raw data for constructing and testing the NNs comes from MQCT calculations, which are computationally expensive, ideally one would like to have the smallest set of MQCT data to train and validate the models. With this in mind, we tested a further reduction of the size of the training and validation data by limiting the state-to-state transitions to $|\Delta E|\le200$ cm$^{-1}$. The resulting NN predictions are shown in Figure S9 and Figure S10 of the Supplementary Material for para and ortho H$_2$O, respectively. The TACS are shown in Figure S11 of the Supplementary Material. Clearly, the agreement is not at the same level as with Dataset 3. Higher discrepancies are visible at the level of individual cross sections as well as TACS but they remain reasonable. We conclude that Dataset 3 provides the best balance between performance and accuracy.
		
\subsection{Computational efficiency of the ML models}
It is clear that the efficiency of the ML approach presented here is dependent on the size of the training and validation dataset since that is the data one needs to compute using MQCT. Therefore, the NNs constructed from Dataset 4 is the most efficient while Dataset 1 is the least efficient. Due to the higher error of the predicted results from Dataset 4, we deem it unacceptable. This leads us to conclude that Dataset 3 provides an optimal choice to achieve acceptable accuracy while keeping the cost of computing the required MQCT data for training and validation reasonably small. Thus, the following analysis pertains to Dataset 3.
		
Besides the size of the training and validation data there are several other approximations that one can make to reduce the computational cost of the proposed ML approach. One can eliminate small-magnitude cross sections to reduce noise, but these cross sections are hard to identify without scattering calculations. Therefore, we do not consider this for our efficiency estimation. Another approximation is made based on the magnitude of the range of $|\Delta E|$. This can be explored prior to large scale MQCT calculations in order to achieve the most efficient workflow proposed in our methodology.
		
On average, the size of the test data is $\sim7.5$ times larger than the sum of the sizes of the training and validation data. It was found from  previous studies by Mandal et al.~\cite{mandal2023adiabatic} that the cost of MQCT calculations, within the adiabatic trajectory approximation (AT-MQCT) methodology,  grows as $N^3$ (typical of coupled-channel calculations), where $N$ is the number  of rotational states  in the basis. Note that a larger rotational basis is needed for higher $|\Delta E|$ transitions. Thus, excluding higher $|\Delta E|$ transitions, as discussed in the various architecture of datasets, can drastically reduce the computation cost.
		
The efficiency of the ML approach should consider both the cost of constructing the relevant transition matrices and the actual cost of propagating the trajectories. However, these matrices now also contain significantly reduced numbers of transitions compared to the original MQCT calculations. In the past, the cost of computation of each of these four matrices for the original MQCT calculations was about $676,000$ CPU hours each and a total of four matrices were computed. However, it is difficult to estimate the cost savings in computing these smaller matrices. Additionally, a newer version of MQCT is expected to significantly speedup these calculations making them computationally less demanding compared to the MQCT trajectory simulations. Thus, we do not include it in our estimation of computational efficiency
		
As reported earlier~\cite{mandal2023rate}, the cost of the MQCT trajectory calculations was about $5,250,000$ millions CPU hours. Considering that the test data for Dataset 3 is about 7 times larger than the training and the validation data and that computationally demanding high $|\Delta E|$ transitions ($|\Delta E|>300$ cm$^{-1}$) were excluded in the dataset, we estimate the cost for the trajectory calculations to be around $100,000$ CPU hours for producing the relevant data for training and validation. Overall, we expect about a factor of 50 savings in computational cost from our approach. The actual CPU time for training these NNs is insignificant (about $\sim 5.5$ CPU hours) once the training and validation data are in place. This is a remarkable gain in efficiency.  Using the methodology presented here, an expanded database of much needed rotational transitions in water molecules can be computed at reasonable computational cost.
							
\section{Conclusion}
\label{sec:conclusion}
In this work, we implemented a machine-learning method to reduce the cost of computing state-to-state and thermally averaged cross sections for collisions of complex molecular systems, such as H$_{2}$O + H$_{2}$O. Due to methodological as well as computational  limitations, inelastic collisions of these systems remain largely unexplored while the rate coefficients are much needed for the astrophysics community. Computational costs of building a database of rate coefficients for rotational transitions in H$_{2}$O + H$_{2}$O collisions were estimated to be $\sim$ $5-8$ million CPU hours when only first 231 and 210 transitions of para- and ortho-H$_{2}$O molecules, respectively, were explored. However, this is not sufficient and in this work we present a methodology to expand the database further to include more rotational transitions at significantly reduced computational cost.
		
Prior applications of GP and NNs for predicting state-to-state cross sections/rate coefficients for atom-diatom scattering involved interpolation in spaces of quantum states defined by only four quantum numbers ($v,j$ and $v',j'$) of the diatomic molecules~\cite{bossion2024machine}. Similarly, in the application of GP models for diatom-diatom scattering, at most 8 quantum numbers ($v_i,j_i$, $v_i',j_i'$ of two diatomic molecules) are considered in the work of Jasinski et al.~\cite{jasinski2020machine} though  recent studies of Mihalik et al.~\cite{mihalik2025accurate} and Wang et al.~\cite{wang2025multilayer} considered only changes in ro-vibrational levels of SiO in collisions of SiO and ground state para-H$_2$. Here, theML approach is shown to yield reliable results for interpolation in the space of 12 quantum numbers for H$_2$O+H$_2$O collisions. Our approach of using an ensemble of NNs for a range of collision energies paves a pathway for efficient computation of rate coefficients for state-to-state rotational transitions in collisions of two asymmetric top molecules.
		
For practical purposes, an estimated speed up by a factor of $\sim$50 is expected using our proposed pipeline exploiting the physics behind the energy transfer process (exponential decay of rate coefficients with energy gap) and utilizing deep learning algorithms. The cross section data for rotationally  inelastic scattering of two H$_{2}$O molecules computed using MQCT shows a double exponential feature as a function of the energy difference between the initial and final states. The MLP algorithm using neural network appears to capture this feature during training and yield a model that successfully predicts cross sections for state-to-state rotational transitions in H$_{2}$O + H$_{2}$O collisions from which accurate thermally averaged cross sections are derived. Our tested NNs achieved an excellent accuracy level for deviations for higher magnitude of thermally averaged cross sections. In the future, we hope to use this methodology to compute additional rotational transitions in water to extend the existing database.
		
The methodology presented here is robust and general, and can be implemented for other systems of complex colliding partners, such as HCN + H$_{2}$O, HDO + H$_{2}$O, and collisions of atoms and diatoms with other polycyclic aromatic hydrocarbons (PAHs). By utilizing the machine learning models using neural networks as proposed in this work, these computationally demanding scattering calculations are expected to become significantly more affordable. This proposed workflow is expected to open a new avenue in the near future to populate databases such as BASECOL for astrophysical modeling.

\section*{Author contributions}
N. B. and B. M. conceived the project. B. M. and D. B. generated the original MQCT data. B. M. constructed and tested the ML models with assistance from all co-authors. All co-authors contributed to data analysis, validation,  and manuscript preparation.
%{\color{red}We strongly encourage authors to include author contributions and recommend using \href{https://casrai.org/credit/}{CRediT} for standardised contribution descriptions. Please refer to our general \href{https://www.rsc.org/journals-books-databases/journal-authors-reviewers/author-responsibilities/}{author guidelines} for more information about authorship.}

\section*{Conflicts of interest}
There are no conflicts to declare.

\section*{Data availability}
The data supporting this article have been included in the main article and the Supplementary Information.

\section*{Acknowledgements}
This work is supported in part by NSF grant No. PHY-2409497 (N. B.) and  NASA grant  80NSSC22K1167 (P.C.S.). D. B. was supported by NSF grant CHE-2102465. R.V.K. acknowledges support from Natural Sciences and Engineering Research Council (NSERC) of Canada. This research also used HPC resources at Marquette University funded in part by NSF award CNS-1828649 and the RebelX cluster at UNLV.

%%%END OF MAIN TEXT%%%


\balance


%%%REFERENCES%%%
\bibliography{rsc} %You need to replace "rsc" on this line with the name of your .bib file
\bibliographystyle{rsc} %the RSC's .bst file

\end{document}